\documentclass[journal]{IEEEtran}
\usepackage{hyperref}  % hyperlinks
\usepackage{amsfonts, amsmath, amssymb, amsthm}
\usepackage{graphicx}
\usepackage{booktabs}
\usepackage[table]{xcolor}

\newtheorem*{remark}{Remark}

\theoremstyle{definition}
\newtheorem{definition}{Definition}

\begin{document}
%MSWord compatible
%% allowing the author to define a "short title" to be used in page headers.TACOF
% Pre-learning/task-agnostic/taskless/+colab.
% Weakly Federated Systems, TACOF
\title{Task-Agnostic Federation over Decentralized Data: Research Landscape and Visions}
% \title{Task-Agnostic Collaboration in Weakly Federated Systems over Decentralized Data: Research Landscape and Visions}
% \title{Knowledge Augmentation in Federation: Rethinking The Value of Collaboration over Decentralized Data}

\author{Wentai~Wu,~\IEEEmembership{Member,~IEEE,}
        Ligang~He*,~\IEEEmembership{Member,~IEEE,}
        Saiqin~Long,
        Ahmed~M.~Abdelmoniem,~\IEEEmembership{Senior Member,~IEEE,}
        Yingliang~Wu,
        Rui~Mao, and
        Keqin Li,~\IEEEmembership{Fellow,~IEEE}% <-this % stops a space
\thanks{W. Wu is with the Department of Computer Science, College of Information Science and Technology, Jinan University, Guangzhou 510632, China. e-mail: wentaiwu@jnu.edu.cn.}% <-this % stops a space
\thanks{L. He is with the Department of Computer Science, University of Warwick, Coventry CV4 7AL, United Kingdom. e-mail: ligang.he@warwick.ac.uk.}% <-this % stops a space
\thanks{S. Long is with the Department of Computer Science, College of Information Science and Technology, Jinan University, Guangzhou 510632, China. e-mail: saiqinlong@jnu.edu.cn.}% <-this % stops a space
\thanks{A. M. Abdelmoniem is with the School of Electronic Engineering and Computer Science, Queen Mary University of London, London E1 4NS, United Kingdom. e-mail: ahmed.sayed@qmul.ac.uk.}% <-this % stops a space
\thanks{Y. Wu is with the Department of Electronic Business and the Institute of Digital Business \& Intelligent Logistics, South China University of Technology, Guangzhou 510006, China. e-mail: bmylwu@scut.edu.cn.}% <-this % stops a space
\thanks{R. Mao is with the College of Computer Science and Software Engineering and the Shenzhen Institute of Computing Sciences, Shenzhen University, Shenzhen 518060, China. e-mail: mao@szu.edu.cn.}% <-this % stops a space
\thanks{K. Li is with the Department of Computer Science, State University of New York, New Paltz, New York 12561, USA. e-mail: lik@newpaltz.edu.}% <-this % stops a space
\thanks{*Corresponding author: Ligang He}
% \thanks{Manuscript received November 19, 2024}
}


\markboth{Preprint under review}%
{Wu \MakeLowercase{\textit{et al.}}: TAF}

\maketitle

\begin{abstract}
Increasing legislation and regulations on private and proprietary information results in scattered data sources also known as the ``data islands''. Although Federated Learning-based paradigms can enable privacy-preserving collaboration over decentralized data, they have inherent deficiencies in fairness, costs and reproducibility because of being learning-centric, which greatly limits the way how participants cooperate with each other. In light of this, we investigate the possibilities to shift from resource-intensive learning to task-agnostic collaboration especially when the participants have no interest in a common goal. We term this new scenario as Task-Agnostic Federation (TAF), and investigate several branches of research that serve as the technical building blocks. These techniques directly or indirectly embrace data-centric approaches that can operate independently of any learning task. In this article, we first describe the system architecture and problem setting for TAF. Then, we present a three-way roadmap and categorize recent studies in three directions: collaborative data expansion, collaborative data refinement, and collective data harmonization in the federation. Further, we highlight several challenges and open questions that deserve more attention from the community. With our investigation, we intend to offer new insights about how autonomic parties with varied motivation can cooperate over decentralized data beyond learning. 
\end{abstract}

% key terms: data-centric, generic knowledge
\begin{IEEEkeywords}
Generic knowledge, Decentralized data, knowledge exchange, data engineering, federated systems.
\end{IEEEkeywords}

% \fi
\IEEEpeerreviewmaketitle

\section{Introduction}
\label{sec:intro}
As Artificial Intelligence (AI) exhibits its potential in transforming human society, we are entering a new era where the amount of data and the value generated from data both increase in an unprecedented speed. Statistics show that the total volume of data or information created, captured, copied and consumed globally has experienced a tenfold growth in the past decade and is expected to reach 394 zetabytes in 2028 \cite{statista2028}. A research predicted that a single autonomous vehicle can generate as much as 10 gigabytes data per second in the future \cite{yang2020vehicle}. This signifies two trends in the development of Machine Learning (ML) and data mining techniques. On the one hand, vast data lays the foundation for large-scale learning systems \cite{guo2024bls,guo2025deepseek,zhong2024evaluation}. On the other hand, knowledge sharing across multiple data sources matters more than ever \cite{zhao2024domain, 
guan2025effective}. 

%Knowledge sharing can alleviate inconsistency and ambiguity of information for classic problems \cite{jia2025instance,hang2025partial}.
In reality, the accessibility of private data is in many cases limited by legislation and regulations such as the General Data Protection Regulation (GDPR)\footnote{https://gdpr-info.eu/} and the California Consumer Privacy Act (CCPA)\footnote{https://www.oag.ca.gov/privacy/ccpa}. This to a great extent leads to a paradigm shift from centralized learning to collaborative training over decentralized data. The most popular solution to ``data islands'' is Federated learning (FL) \cite{mcmahan2016arxiv, mcmahan2017communication}, which was developed on the basis of Federated Optimization for collaborative model training \cite{kairouz2021fl}. %Essentially, FL is characterized by massive system scale, restricted communication and most importantly, the non-Independent and Identically Distributed (non-IID) data assumption \cite{kairouz2021fl}. 
Although extensive researches have been conducted following the FL paradigm, \emph{they hardly go beyond learning-centric, task-specific knowledge sharing}. This protects the raw data but results in several deficiencies:
\begin{itemize}
    \item \textbf{Intractable fairness}: No matter the participants cooperate towards a consensus model for common welfare or personalized models for individual interest, practical fairness is hard to realize. This is because the payoff and cost differ greatly between the parties given pathological data distribution, but learning-centric paradigms such as FL generally struggle to balance the two.
    \item \textbf{Prohibitive costs}: Models are ``heavy'', and so are the model updates and the gradients. They are generally in the same tensor structure and can take up to gigabytes of network traffic for each transmission. Besides, the cost for local training cannot be overlooked considering the problem of slow convergence over non-IID data \cite{amiri2022conv}.
    \item \textbf{Poor reproducibility}: models can carry knowledge but are hard to reproduce in massive collaboration due to the randomness in participation. Any participant has to request massive cooperation in order to reproduce a model from scratch. 
\end{itemize}

\begin{figure*}[htb!]
    \centering
    \includegraphics[width=0.92\linewidth]{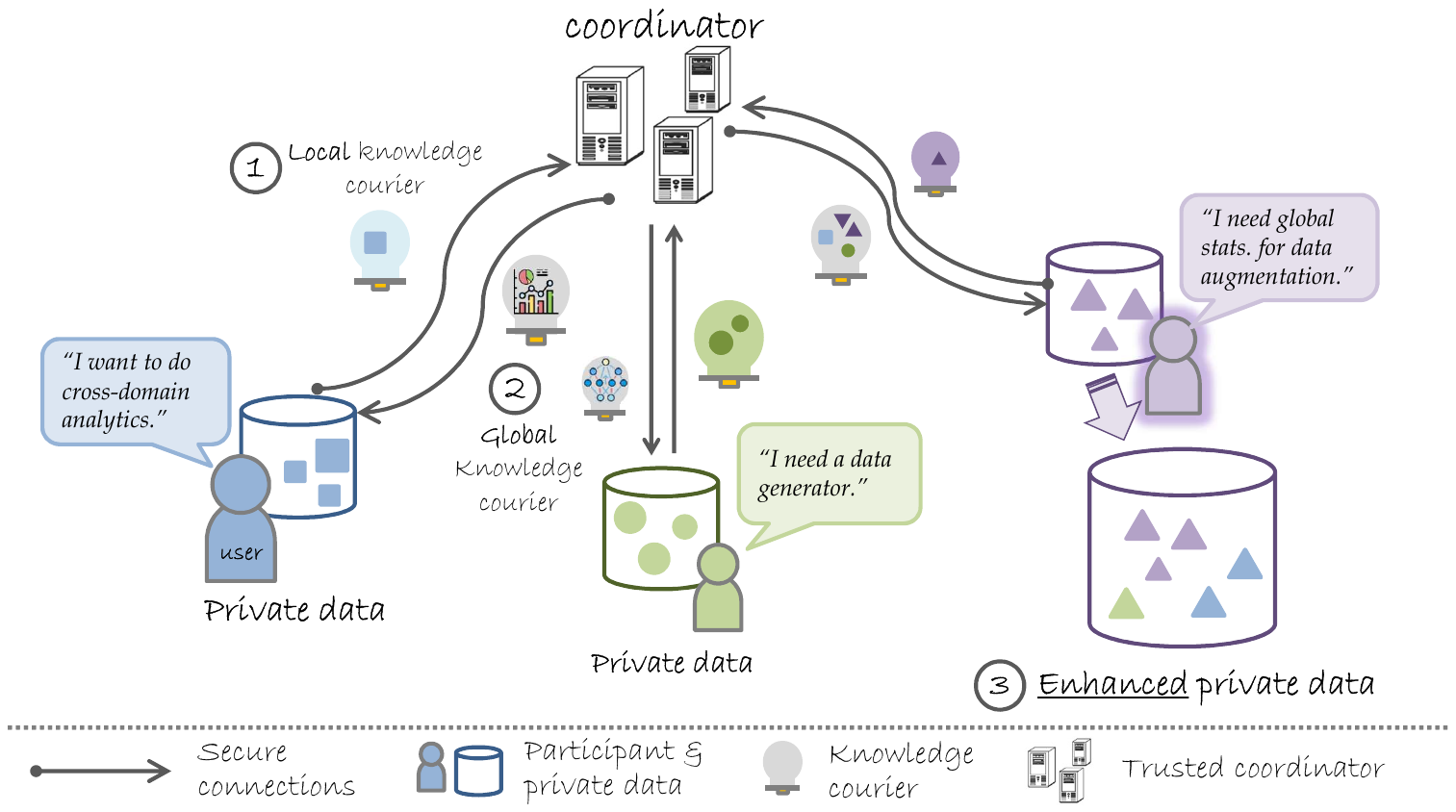}
    \caption{The conceptual overview of Task-Agnostic Federation (TAF), which is a data-centric scenario that liberates the participants from any common goal and employs flexible protocols for generic knowledge exchange.}
    \label{fig:overview}
\end{figure*}

\emph{Is it possible to find an alternative to task-oriented collaboration and tackle the above-mentioned issues?} In this paper we offer our answer by describing \textbf{a new scenario termed Task-Agnostic Federation (TAF), under which we gather emerging studies that enable new ways of collaboration through the exchange of generic knowledge in data}. TAF liberates the participants by allowing for diverse motivation of collaboration in a federated system. This is done by shifting the focus from learning models to enhancing data across the domains, as illustrated in Fig. \ref{fig:overview}. This generalizes the benefits of collaboration because the enhanced data can take various forms and serve for any downstream purposes beyond learning. %Technically, KAF makes difference through i) the introduction of data-centric objectives for collaborative optimization in the federated system, and ii) the adoption of more flexible carriers of knowledge without compromising privacy or proprietary rights. 
The concept of TAF will be elaborated in Sec. \ref{sec:overview}.

\begin{figure}[htb!]
    \centering
    \includegraphics[width=0.99\linewidth]{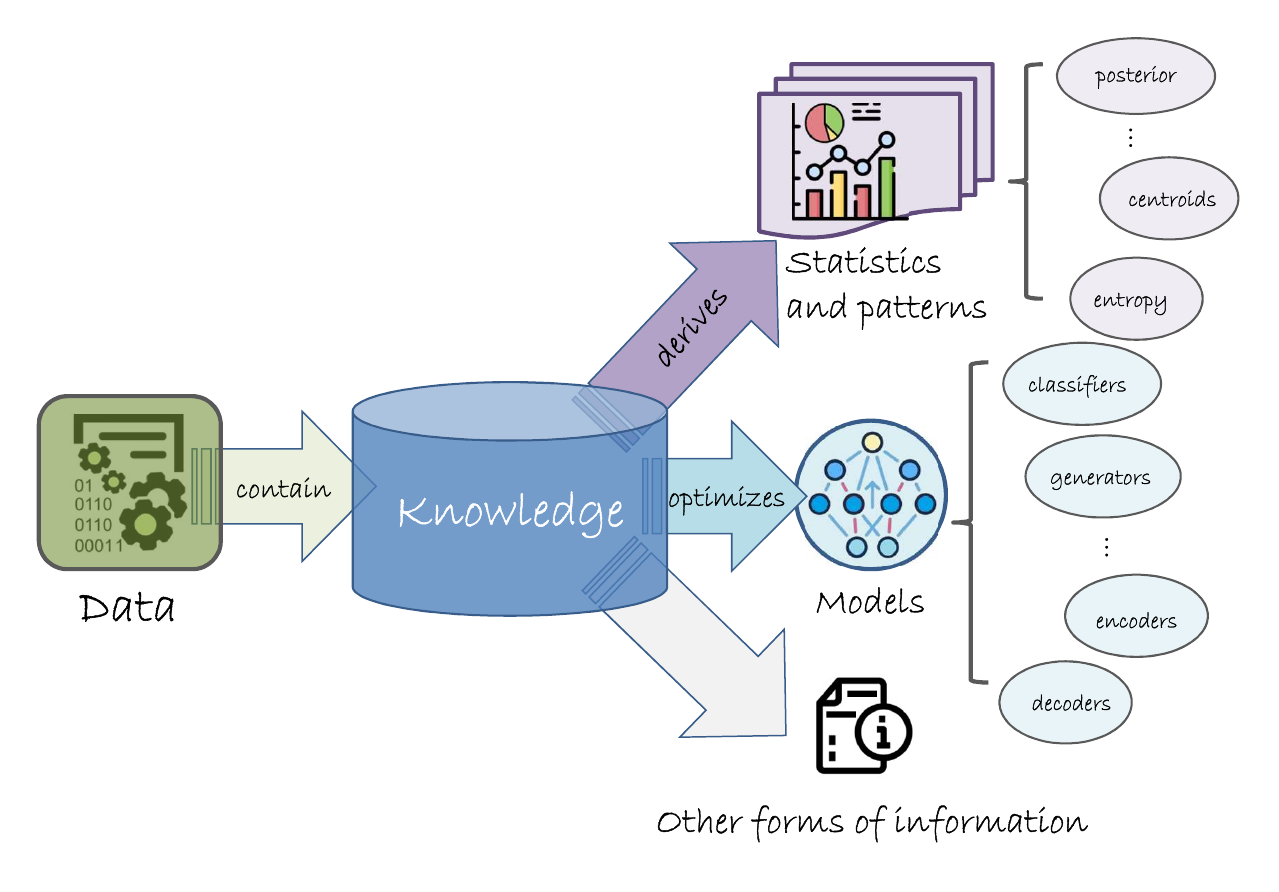}
    \caption{Data is the raw material and the most flexible form of knowledge. Information presented in statistics, patterns and models is derived from data and carries a specific subset of knowledge.}
    \label{fig:knowledge_venn}
\end{figure}

Traditional distributed learning paradigms seek to build a strong model by aggregating models, gradients or other forms of knowledge from separate domains. Even in heterogeneity-aware approaches such as personalized Federated Learning (pFL) and Federated Multi-Task Learning (FMTL) \cite{rahman2024tvt, liu2025pfed, skovajsova2025fmtl}, the optimization is still mostly limited to task-specific models which are hardly generalizable. By contrast, Task-Agnostic Federation emphasizes the enhancement of private data with generic knowledge across the system. Particularly, The knowledge brought in can take whatever form, e.g, statistical information, synthetic data, generators, class prototypes, predictions, etc., as long as it serves to enrich or refine the knowledge implied in local data (Fig. \ref{fig:knowledge_venn}). This calls for new protocols that enable cost-effective exchange of knowledge under privacy constraints. Compared to existing paradigms, TAF has advantages in several aspects. First, it directly enhances local data in quantity, quality or utility. Second, it brings generic, reusable knowledge to users that can be utilized later for diverse custom purposes. Besides, it leads to sustainable gain for every participating party despite how much they are engaged\footnote{By contrast, FL relies on persistent client engagement to produce strong models.}. This is practically important for the participants who have very limited resources to invest. 

% massive data, learning, privacy(GDPR, CPCD, etc.)
% data islands, FL, key=model-based knowledge sharing, limit=fairness+cost+reproduce
% Alternative? i) carrier: data=observable form of knowledge, model=an estimate of knowledge, ii) objective of collaboration
% kaf = fka
% advantages:
% 1. Improve global knowledge coherence
% 2. Enhance local data in quantity, quality and utility for better local training
% 3. data-centric reproduce

\textbf{In fact, an increasing number of studies have adopted inspiring methodologies in relevance to TAF, but a comprehensive review is so far missing in the literature.} As far as we know, this review is the first to assemble related studies in the field under this concept.

% ours
\noindent\emph{\textbf{Our contributions.}} In this paper we advocate a data-centric vision and focus on how multiple parties can cooperate without a common goal in a federated system. We believe TAF distinguishes itself as a new roadmap for autonomic intelligence, where each participant seeks and shares knowledge that is generic and reusable. The value of this article is summarized as follows:
\begin{itemize}
    \item A revisit to multi-party collaboration across ``data islands'' from the standpoint of task-irrelevant interest and long-lasting benefits.
    \item New understanding of collaboration as a bridge to augmenting, refining and transforming local data for general purposes rather than learning.
    \item Literature review and insights for what TAF can bring up for the new forms of federated systems where autonomy matters.
\end{itemize} 

% survey comparison
\begin{table*}[tb!]
    \centering
    \caption{The comparison between our work and the surveys in related fields regarding the research themes.}
    \begin{tabular}{l p{3.5cm} c c c c}
    \toprule
    Theme & Surveys & Task-agnostic & Collaboration & Learning & Local benefit \\\addlinespace[1pt]
    \midrule
    Data engineering \& knowledge management & (e.g., \cite{balayn2021biasunfairness,cui2025kadl,xie2024dataquality,gonzales2023synth,caton2024fairness,mehrabi2022biasfairness}) 
    & \checkmark  &  &  & \checkmark\\\addlinespace[2pt]
    
    Federated learning and variants & (e.g.,  \cite{kairouz2021fl,chen2024flprivacy,martinez2023decenfl,sabah2024model,ye2023heterofl,ji2024fxl,skovajsova2025fmtl})
    &  & \checkmark & \checkmark & \checkmark (pFL\&FMTL)\\\addlinespace[2pt]

    Federated analytics & (e.g., \cite{wang2025fa,wang2022fa,elkordy2023federated})
    &  & \checkmark &  &  \\\addlinespace[3pt]
    
    \rowcolor{cyan!11}\textbf{Ours} &  & \checkmark & \checkmark & \checkmark & \checkmark \\\addlinespace[2pt]
    \bottomrule
    \end{tabular}
    \label{tab:surveys}
\end{table*}

\section{Background and Related Surveys}
In this section we briefly review some closely related concepts and draw comparison with related surveys to highlight our unique perspectives.
% FL(client), imbalance+re-balancing(beta, inverse samp.), data aug (SMOTE, ADAptive SYNthetic (ADASYN), mixup, GAN), feature engineering

% fed. data engineering
\noindent\textbf{Data engineering and knowledge management.} Data engineering is something usually ``behind the screen'' but vital to any learning or mining pipeline \cite{balayn2021biasunfairness}, whilst the management of knowledge is pervasive in practice spanning from data abstraction to expertise transfer. The rise of these topics dates back to decades ago when rule-based symbolism and expert systems prevailed. An early work by Studer et al. \cite{studer1998knowledge} signified a shift towards modeling view and outlined two diverged approaches to refining knowledge, whilst a recent survey by Cui et al. \cite{cui2025kadl} reviews how deep learning can be augmented by domain knowledge. A diversity of data-centric studies were also investigated in the literature, with topics covering data balancing \cite{xie2024dataquality, gonzales2023synth, chawla2002smote}, feature engineering \cite{chandrashekar2014survey, salau2019extract} as well as data bias and fairness issues \cite{caton2024fairness,balayn2021biasunfairness,mehrabi2022biasfairness}. However, new scenarios like collaborative data engineering and decentralized knowledge augmentation is yet to be covered in existing surveys.

% FL
\noindent\textbf{Federated learning and its variants.} Federated learning is a distributed machine learning paradigm for privacy-preserving collaboration across a multitude of devices \cite{bonawitz2019flsys}. 
% two-layer distribution
The most unique challenge for FL is the multi-party ownership of data. %This significantly complicates the way we operate across multiple disjoint sets of data, i.e., ``data islands''. %On the one hand, in the FL setting we need to deal with a second layer of data distribution, that is, a distribution of similar but different domains represented by client data. On the other hand, we still face the notorious problems, such as data imbalance, sample corruption and label noise, probably within each of the local datasets. More importantly, the orchestrator (e.g., a central server in most cases) is typically agnostic of local data distribution due to privacy constraints.
Kairouz et al. \cite{kairouz2021fl} presented one of the most comprehensive surveys on FL, covering the efficiency, effectiveness, privacy, robustness and fairness issues. %They also highlighted the challenges related to ill-distributed data for both cross-device and cross-silo FL scenarios.
Some surveys focus on particular aspects of an FL system such as privacy preservation \cite{chen2024flprivacy}, decentralization \cite{martinez2023decenfl}, personalized FL (pFL) \cite{sabah2024model}, and heterogeneous FL \cite{ye2023heterofl}. Application-specific studies are also reviewed in the literature, e.g., FL for Internet of Things (IoT) \cite{nguyen2021fliot}, smart cities \cite{pandya2023flcities}, and healthcare \cite{taha2023flhealthcare}. Ji et al. \cite{ji2024fxl} termed the variants of FL paradigms as ``Federated X Learning'', which represents the fusion of FL with other emerging ML paradigms. Nonetheless, these surveys, without any exception, overlooked what learning can bring back to the data.

% FL
\noindent\textbf{Federated Analytics.} Federated analytics (FA) \cite{googleai2020federated}, usually developed on information seeking, confidential computing and cryptography, aims to obtain global statistics with minimal private information leakage. The motivation and unique challenges for FA are discussed in \cite{wang2022fa}, while Wang et al. \cite{wang2025fa} provide a comprehensive view of research landscape, with a taxonomy that entails analytical tasks, algorithms as well as privatization. Elkordy et al. \cite{elkordy2023federated} review relevant techniques to FA from a query-oriented perspective. Yet, FA does not bring knowledge back to the data owners, making what we are interested in uncovered in these surveys.

In contrast to existing surveys, we advocate a task-agnostic vision and investigate how multiple parties in a federated system can benefit from data-centric cooperation. Table \ref{tab:surveys} highlights the aspects that distinguish this review paper from the rest in the literature.

\section{Conceptual Overview}
\label{sec:overview}
In this section we provide a holistic view of the proposed scenario of TAF, starting with a general system architecture and moving on to the problem setting. This is followed by the outline of key approaches that will be elaborated in Secs. \ref{sec:expand}, \ref{sec:filter} and \ref{sec:correct}.

% system, Obj. function, structure

\subsection{System Architecture}
As shown in Fig. \ref{fig:overview}, the system model in TAF is similar to that of (horizontal) federated learning. Considering the diversity of data sources, a central server is usually needed for coordinating the scope and protocol of cooperation. On this hub-and-spoke topology, a multitude of participants $\mathcal{U}=\{u_1,u_2,\ldots\}$ connect to the server to send and receive shareable knowledge. Theoretically, TAF can also employ Peer-to-Peer (P2P) decentralized topology if the community is self-organized. In both cases, each participant $u_k$ has the exclusive access to its local data $D_k$ that is not allowed to be shared for privacy or proprietary reasons. Meanwhile, a certain knowledge exchange protocol has to be established between the participating parties in the first place.   

\begin{remark}
    \textbf{A key feature of TAF is that any participant is not necessarily capable of or interested in training}. This is because generic knowledge does not always come from model training. For example, participants can swap data statistics, feature rankings, or synthetic samples for their specific purposes.
\end{remark}

\subsection{Formal Definition}
The fundamental setting of TAF determines that it is established over ``data islands''. This means that any participant $u_k$ in the federation $\mathcal{U}$ is associated with a private dataset $D_k$. Each participant $u_k$ exchanges messages with the coordinator by a customized protocol $\pi_k$, which determines what it needs to send and receive and how. The formal definition is given as follows:

\begin{definition}[Task-Agnostic Federation (TAF)]
\label{def:taf}
    \textit{Involving a set of autonomic participants $\mathcal{U}$ each with a private dataset $D_k$, TAF is a collaborative system or scenario where each participant $u_k$ seeks to maximize its data-related gain $g_k(D_k, \mathcal{U}, \pi_k)$, where $g_k$ is a participant-specific function reflecting how much $D_k$ is enhanced. Each participant operates by a custom protocol $\pi_k$ without knowing any other participant's objective.}
\end{definition}

\noindent\textbf{Example.} To make the objective more understandable, here we sketch a mock algorithm. First, we use the \textit{effective number of samples} proposed in \cite{cui2019cbloss} to define $\nu(D)\triangleq (1-\beta^{|D|})/(1-\beta)$, with $\beta$ set as a priori (e.g., $\beta=0.999$). Next, each participant performs data clustering, and reports its centroids to the server in exchange for other participants' centroids following the protocol $\pi_k$\footnote{We are not concerned about privacy leakage in this toy example.}. The shared centroids are then used for some local data augmentation $\Theta_{\mathcal{U}, \pi_k}: D_k \rightarrow D'_k$ (e.g., interpolation), which translates to $u_k$'s gain expressed as $g_k(D_k, \mathcal{U}, \pi_k) = \Delta \nu = \nu(D'_k) - \nu(D_k)$. %Finally, the clients compute the \textit{enhancement} and \textit{regularization} terms while the server computes the \textit{alignment term} in (\ref{eq:obj1}), with some centroid-based distance measure $d(\cdot,\cdot)$.
The gain increases as the algorithm iterates.

% Collectively, one can derive a system-wide gain from (\ref{def:taf}) if all the participants are cooperative and stick to their protocols $\{\pi_1,\pi_2,\ldots\}$:
% \begin{equation}
%     \max_{\{\pi_1,\pi_2,\ldots\}} \sum_{k=1}^{|\mathcal{K}|} F_k.
%     \label{eq:obj2}
% \end{equation}

\begin{remark}
    The gain implies that each participant in TAF is self-motivated and remains agnostic of any task (including its own and other participants'). On this point, TAF differs from pFL and FMTL as they aim to minimize a set of task-specific losses w.r.t. a set of specific models. Further, TAF brings sustainable benefit to the participants since their local data, once enhanced, can serve any downstream purposes beyond training.
\end{remark}

\begin{figure*}[htb!]
    \centering
    \includegraphics[width=0.96\linewidth]{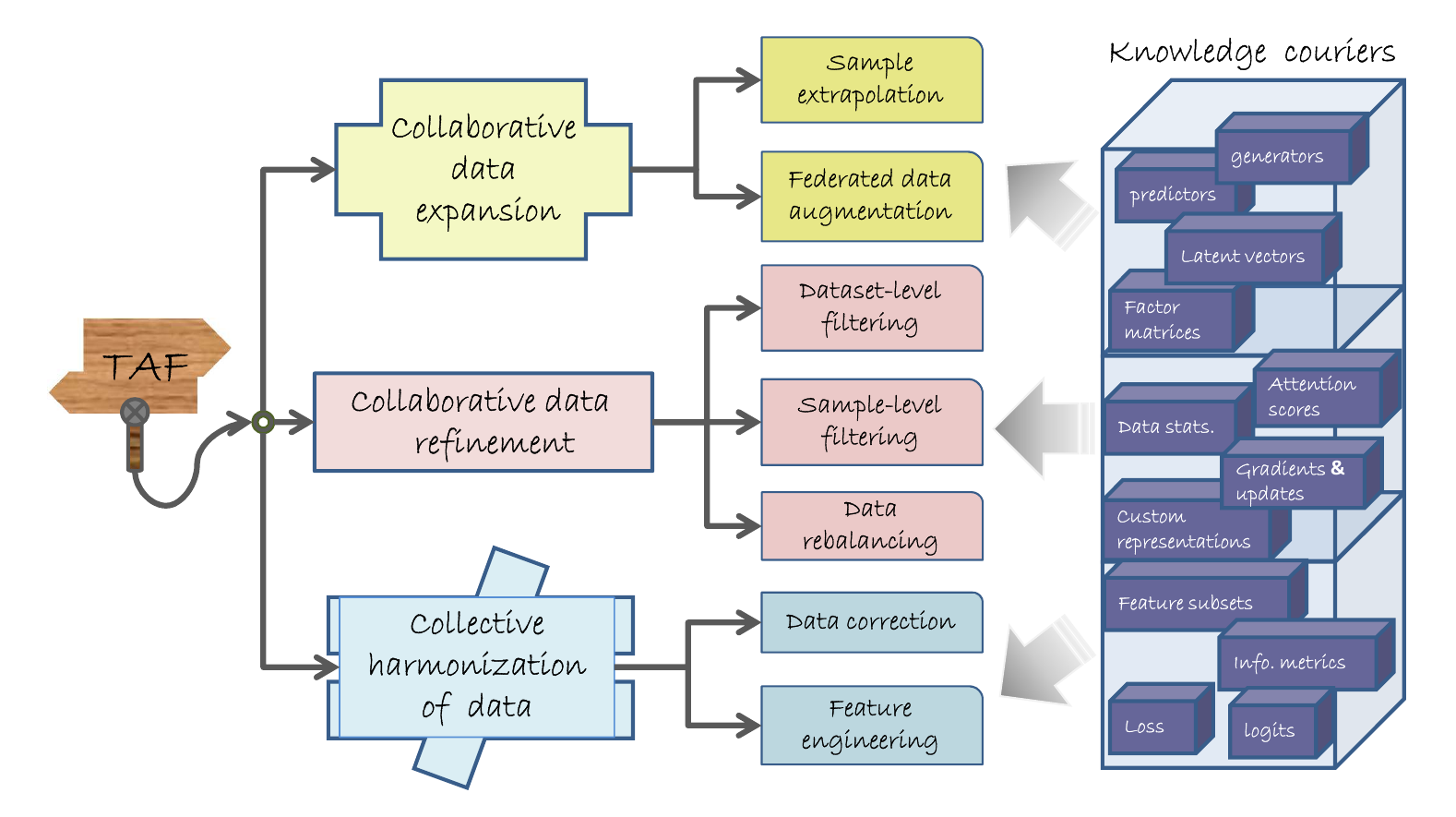}
    \caption{(Left) Taxonomy of research related to task-agnostic federation, and (right) the couriers of knowledge employed in the studies.}
    \label{fig:struct}
\end{figure*}

\subsection{Outline of Approaches}
Through an in-depth investigation, we collect a rich set of studies that align or partially align with the characterized scenario of TAF. Some of these papers rest in the scope of federated learning but employ data-centric methods suitable for task-agnostic scenarios. We adopt a two-layer taxonomy: in the top layer we define three types of data-centric collaboration: collaborative data expansion, collaborative data refinement, and collective harmonization of data, and in the bottom layer we group the papers into multiple branches based on their technical design. Fig. \ref{fig:struct} visualizes the taxonomy of studies covered by this survey.

\section{Collaborative Data Expansion} % Addition
\label{sec:expand}
%As discussed previously, a federated system entails a set of knowledge domains that collectively comprise the global domain. Intuitively, it is easy to understand that greater overlaps between the local domains can facilitate faster and better learning. %For example, it is commonly agreed that the global model's convergence can be improved by reducing the ``non-IIDness'' of local data. In fact, this can be achieved by outspreading the knowledge contained in each local domain.

In this section, we investigate popular approaches for collaborative data expansion under the federated setting. By expansion we refer to methods that outspread the knowledge contained in each local domain by diversifying data samples.%where the implications of expansion are two-fold. First, this represents a roadmap for extending the boundaries of local knowledge domains through introducing exogenous information from other clients. Second, it also refers to the augmentation of data with the aims of re-balancing local distribution and generating diversified instances. 
% Hence, the key criteria for this research category are:
% \begin{itemize}
%     \item Seeking to extend local domains of knowledge through collaboration;
%     \item Improving the diversity of data samples within each local domain based on information exchange.
% \end{itemize}

We summarize that there are two major lines of research in this direction. In the following content, we first introduce existing studies for extrapolating the samples in local data by information sharing and reconstruction. Then we revisit the research on enhancing local data in terms of size, quality and distribution via data augmentation.

\subsection{Federated Sample Extrapolation}
\begin{figure}[htb!]
    \centering
    \includegraphics[width=0.99\linewidth]{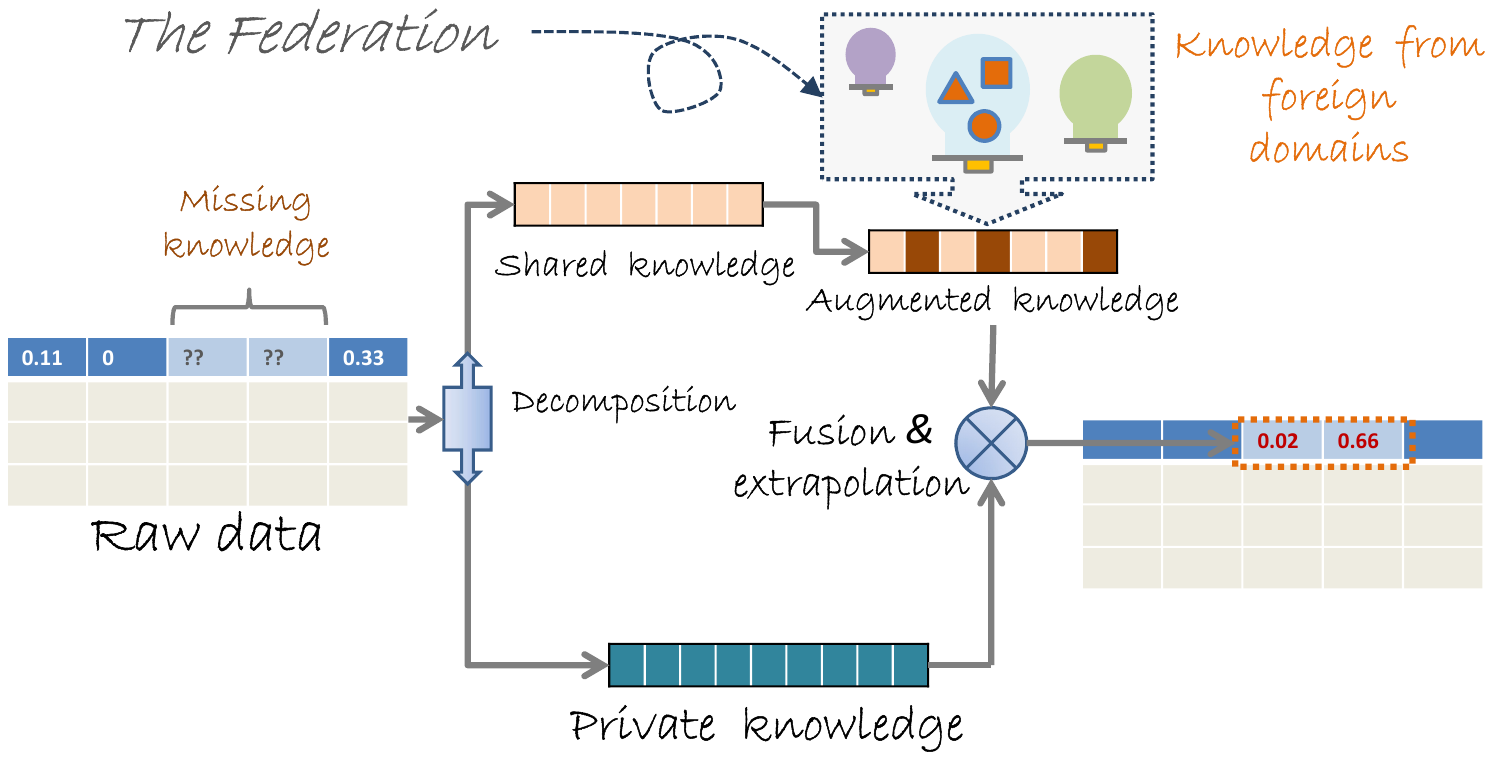}
    \caption{A schematic showing the general framework of federated sample extrapolation. The goal is to obtain the missing part of knowledge in the local domain.}
    \label{fig:know_extrapolate}
\end{figure}

%Fed. Collaborative filtering
We conceptually describe sample extrapolation as the methodology of filling the missing part of information that is not explicitly expressed by the raw data. This process usually involves predicting or approximating implicit samples locally by utilizing some global knowledge contributed and shared by other participants in the system (Fig. \ref{fig:know_extrapolate}). In a federation, sample extrapolation can be very useful when all or most of the local datasets are sparse in terms of information density. For example, Sanyal et al. \cite{Sanyal2019fff} considered a healthcare scenario where numerous Internet of Medical Things (IoMT) devices act simultaneously as independent data collectors. The challenge here is that their readings need to be collectively utilized for analysis, but it is not encouraged to continuously push them to a central server. To this end, the authors proposed a Federated Filtering Framework (FFF) \cite{Sanyal2019fff}, where the key component on each local device is a predictor for approximating the true sensing data. The IoMT devices communicate with a fog server that aggregates and utilizes the predictors in order to reconstruct the full data matrix. %For communication efficiency, the timing of local model update and aggregation is determined by dynamic thresholds. 

Sparsity is one of the major obstacles that can largely undermine the efficacy of machine learning and data mining. The problem has been extensively explored in the context of recommender systems where the data are represented as a user-item interaction matrix that is huge in dimensionality but low in information density, which is caused by the natural pattern of user behavior and the scale of e-commerce platforms. Although there are many solutions to it in the literature, most of the existing studies make the assumption that the whole data matrix is available for analysis. However, a centralized collection of user behavior data may not be permitted in many situations such as cross-border e-commerce. 

As one of the most popular approaches to recommender systems, {Collaborative Filtering (CF) can be viewed as a data model that is able to extrapolate incomplete information (i.e., unrecorded interactions) using the dense representation of factor matrices.} Approved by the Netflix Prize, Matrix Factorization (MF) is the classic implementation of CF, where the raw user-item interaction matrix $\boldsymbol{R}\in\mathbb{R}^{m\times n}$ can be approximated as
\begin{equation}
    \boldsymbol{R} \sim \boldsymbol{X}^T \boldsymbol{Y},
\end{equation}
where $\boldsymbol{X} \in \mathbb{R}^{k\times m}$ and $\boldsymbol{Y} \in \mathbb{R}^{k\times n}$, with $m$ and $n$ being the number of users and items, respectively.

% Fed. CF (MF)
It is hard to predict unknown interactions (i.e., new knowledge) without a centralized storage of user vectors (which are likely to contain user-private information). Ammad-ud-din et al. \cite{FCF2019} first described this new model as Federated Collaborative Filtering (FCF), where all the participants keep their interaction data private. Based on a reconstruction loss and stochastic gradient descent (SGD), participants in FCF update their user-specific (user-factor) matrices locally given the latest item-factor matrix from the server. In the meantime, each participant also calculates its corresponding part of the item-specific gradients and feeds back to the server for item-factor matrix update. Following a similar framework, Yang et al. \cite{yang2024discrete} further investigated a multi-behavior scenario, where both item viewing and purchase are recorded but the former is kept private by each user. The authors proposed to binarize both factor matrices and utilized a caching mechanism to reduce communication payload for collaborative parameter update. Considering that a single form of information may not be sufficient for accurate recommendation, the information provided by implicit and explicit feedback can be integrated using a probabilistic model \cite{yang2022dpmf}.

% Fed. NCF
Neural Collaborative Filtering (NCF) \cite{NCF2017} replaces the matrix factorization process of CF by using deep neural networks (DNNs) to learn the user embedding vectors $\boldsymbol{X}=\{\boldsymbol{x_1}, \boldsymbol{x_2}, \ldots \}$ and item embedding vectors $\boldsymbol{Y}=\{\boldsymbol{y_1}, \boldsymbol{y_2}, \ldots \}$. \emph{NCF differs from MF-based approaches in the sense that knowledge is encapsulated in deep neural networks (DNNs)} rather than factor matrices. Inspired by this property, Perifanis and Efraimidis \cite{perifanis2022federated} designed a federated NCF framework that extends the basic NeuMF model, where each participant holds three sets of parameters including their user profile, item profile, and the DNN's weights.
With user profiles kept private, they proposed a secure aggregation algorithm, MF-SecAgg, to perform server-side parameter update, with separate steps for item profile update and DNN update respectively. With a similar motivation, Jiang et al. \cite{JiangFedNCF2022} introduced an adaptive Differential Privacy (DP) mechanism with decaying noise into the communication process of Federated NCF. This protects the exchange of gradients against inference attacks. 

For effective sample extrapolation, it is intricate to strike a balance between preserving learned representations locally and leveraging the common knowledge shared globally. To address this, Li et al. \cite{li2024personalized} designed a gating dual-encoder structure based on Variational Auto-Encoder (VAE) for federated CF. They introduced a gating network followed by two encoders, one for pertaining local knowledge and the other for learning user representations in the global latent space. By mapping the user-item interaction vectors to separate latent subspaces, the loss of personalized knowledge (due to federated aggregation) is effectively mitigated.

Generally, user-private knowledge needs to be stored locally, no matter as a factor matrix (for MF-based methods) or a neural network's parameters (for NCF). This facilitates global knowledge fusion but also increases the burden on end devices. So is it possible to make local knowledge ``invisible''? Singhal et al. \cite{singhal2021federated} offered their answer by allowing stateless participants to rebuild their local model every time they join the federation with the knowledge provided by the global model. This enables flexible collaboration at scale and, more importantly, indicates that personalized knowledge can be extrapolated by leveraging globally shared knowledge.

\subsection{Federated Data Augmentation}
Data augmentation is a powerful technique for diversifying the data when deployed in a federated system. The value of federated data augmentation is two-fold. On the one hand, it directly introduces new knowledge---typically represented by synthetic samples---to the set of local data. On the other hand, multiple participants can collaboratively perform data augmentation, which brings higher consistency of data distribution across the parties in the federation.

Data augmentation generally involves applying transforms on existing samples or synthesizing new samples based on a priori. Note that the synthesis of data and the training of downstream models can be decoupled, hence a simple implementation is to perform data augmentation in the local pre-processing stage without interleaving with the following learning process. Kundu et al. \cite{Kundu2024monkeypox} followed this approach and developed a federated monkeypox disease detection model on the basis of offline image augmentation. Specifically, a Cycle GAN is pre-trained and deployed locally to generate additional samples to enhance local training. 
Similarly, Ullah et al. \cite{Ullah2024xray} developed an FL-based smart healthcare system, where they employed offline oversampling in order to address the class imbalance of chest X-ray imaging data. More specifically, SMOTE is adopted to mitigate the scarcity of minority-class samples such as those with ``COVID-19'' and ``Viral'' labels \cite{Chowdhury2020covid19}. %One potential limit is that unreasonable samples could be generated without using domain-specific knowledge as constraints. 
Despite being easy-to-implement, offline data augmentation is agnostic of any global knowledge and thus may not serve well for the downstream tasks. The potential limit of offline data augmentation can be explained by considering the boundary of knowledge for each domain, wherein offline data augmentation is essentially analogous to linear and non-linear interpolation within the domain. By contrast, through introducing exogenous information for local data augmentation (i.e., collaborative data augmentation), more generalizable knowledge can be obtained to expand the domain, leading to higher gains for the downstream tasks.

% collab aug.
Although the collaboration of different parties naturally provides a rich set of knowledge sources for data augmentation, how to share the knowledge without compromising user privacy remains the key challenge. Essentially, a participant needs to perceive the distribution of data outside its domain without raw data exchange. On this point, sharing a generator (i.e., a generative model that produces new samples) has proved to be an useful mechanism. Wen et al. \cite{Wen2022CVAE} proposed a generative model-based federated data augmentation strategy termed FedDA. They deployed and collaboratively trained a Conditional AutoEncoder (CVAE) on each participant. Instead of sharing the whole generator, they extracted the decoder's hidden-layer features as the shared knowledge and introduced a knowledge distillation loss to guide the training. This effectively reduces communication overheads. For privacy preservation, noise samples are mixed into each local dataset. 

Intuitively, one can adapt traditional augmentation, such as the popular MixUp \cite{Zhang2018mixup} strategy, to the TAF setting. Given a loss criterion $\ell$ and a model $h_{\theta}$, na\"ive Mixup globally yields the following loss over any generated samples $(\boldsymbol{x}', y')$: 
\begin{equation}
    \ell_{rawmix}(h_{\theta}(\boldsymbol{x}'), y') = \ell\Big(h_{\theta}\big((1-\lambda)\boldsymbol{x}_i + \lambda\boldsymbol{x}_j\big), (1-\lambda)y_i + \lambda y_j\Big),
\label{eq:globmixup_loss}
\end{equation}
where the raw data from two participants $i$ and $j$ are involved, posing privacy issues. %However, the difficulty here is that raw samples need to be shared if we want to mix up local and global knowledge. 

To achieve privacy-preserving augmentation, Yoon et al. \cite{yoon2021fedmix} presented Mean Augmented Federated Learning (MAFL), a simple framework that enables participants to exchange the average of local samples (called ``mashed'' data) along with model updates. Further, through approximating the global MixUp loss (Eq. \ref{eq:globmixup_loss}) with Taylor expansion, they derived that the approximation can be done using local samples and the averaged samples from other participants (Eq. (\ref{eq:fedmix_loss})):
\begin{equation}
    \ell_{FedMix} = \mu \ell\big( h_\theta (\mu \boldsymbol{x}_i), y_i \big) + \lambda \ell\big( h_\theta (\mu \boldsymbol{x}_i), \bar{y}_j \big) + \lambda\frac{\partial\ell}{\partial \boldsymbol{x}} \bar{\boldsymbol{x}}_j,
\label{eq:fedmix_loss}
\end{equation}
where $\mu = 1-\lambda$, $\bar{\boldsymbol{x}}_j=\frac{1}{|\xi|}\sum_{j\in \xi}\boldsymbol{x}_j$ and $\bar{y}_j=\frac{1}{|\xi|}\sum_{j\in \xi}y_j$ represent the mean of features and labels over a private data batch $\xi$ from other clients, respectively. The derivative $\frac{\partial\ell}{\partial \boldsymbol{x}}$ is evaluated at $\boldsymbol{x}=\mu \boldsymbol{x}_i$ and $y=y_i$.

This motivated them to devise the FedMix \cite{yoon2021fedmix} algorithm that performs collaborative MixUp augmentation and training based on a decomposable loss function. Zhang et al. \cite{Zhang2023fedm} further extended this framework for regression tasks using bilateral neighborhood expansion and evaluated it on various public datasets. Empirical results show that their algorithms yielded stronger performance than FedMix \cite{yoon2021fedmix}.

A special case of federated data augmentation is the reconstruction of samples using knowledge from multiple domains, which has proved very useful for accelerated medical imaging \cite{yan2024cross,ahmed2025fedgraphmri}. To adapt to the disparity of imaging operators between different sites, Elmas et al. \cite{elmas2023mri} deployed an unconditional GAN at each site to generate synthetic high-quality samples, with a private discriminator to fit site-specific knowledge and a globally shared generator for prior knowledge embedding. %A tailored reconstruction loss was used to guide the training of local generators.

\begin{table*}[htb!]
    \centering
    \caption{Summarizing the studies related to collaborative data expansion in terms of methodological highlights, drawbacks, datasets for experiments, and the form of knowledge for exchange.}
    \begin{tabular}{p{0.3in}p{0.8in}p{0.7in}p{1.5in}p{1.3in}p{1.1in}}
        \toprule
        Refs. & Category & \textbf{Knowledge carrier} & Highlights & Drawbacks & Datasets\\
        \midrule
         \rowcolor{gray!11}\cite{Sanyal2019fff} & Fed. sample extrapolation & predictors \& filters & $\bullet$ remote data reconstruction via predictors\newline$\bullet$ dynamic communication frequency based on error monitoring & $\bullet$ uniform filter parameters across all devices & MHEALTH\\\addlinespace
         \cite{FCF2019,yang2022dpmf} & Fed. sample extrapolation & factor matrices & $\bullet$ decoupling private and shareable knowledge via matrix factorization & $\bullet$ linear growth of communication cost with the raw matrix's dimensions & MovieLens, FilmTrust, Epinions, and in-house data\\
         \rowcolor{gray!11}\cite{yang2024discrete} & Fed. sample extrapolation & binarized matrices & $\bullet$ finer categorization of private knowledge\newline $\bullet$ binarization and caching & $\bullet$ limited to CF applications & JD, Tmall, User Behavior, and MovieLens\\\addlinespace
         \cite{perifanis2022federated,JiangFedNCF2022,li2024personalized} & Fed. sample extrapolation & latent vectors and prediction models& $\bullet$ better protection of private knowledge\newline$\bullet$ dilution of knowledge in model aggregation due to sparsity & $\bullet$ extra costs for local encoding and model transmission\newline$\bullet$ knowledge redundancy in the shared latent vectors & MovieLens, Lastfm, Foursquare NYC, Ta-Feng, Amazon-Instant-Video, and QB-article\\
         \rowcolor{gray!11}\cite{singhal2021federated} & Fed. sample extrapolation & the global model & $\bullet$ model-agnostic and stateless collaboration\newline $\bullet$ generalizable to diverse tasks & $\bullet$ lack of consideration of data heterogeneity & MovieLens and Stack Overflow\\
         \addlinespace
         \cite{Kundu2024monkeypox,Ullah2024xray}& Federated data augmentation & task-specific models & $\bullet$ easy to implement\newline$\bullet$ low risk of privacy leakage & $\bullet$ only local knowledge exploited in augmentation\newline $\bullet$ hard to generalize & COVID-19 radiography, pox virus, and chest X-ray images\\\addlinespace
         \rowcolor{gray!11}\cite{Wen2022CVAE}& Federated data augmentation & attention scores from generative models & $\bullet$ CVAE for local augmentation\newline$\bullet$ light-weight knowledge exchange & $\bullet$ high local training cost & Fashion-MNIST and CIFAR-10 \\\addlinespace
         \cite{yoon2021fedmix,Zhang2023fedm}& Federated data augmentation & the average of local samples & $\bullet$ controllable anonymity\newline$\bullet$ low communication costs & $\bullet$ hard to generalize for being loss-dependent & FEMNIST, FMNIST, CIFAR-10, CIFAR-100, CINIC-10, Airfoil, NO2, and UTKFace\\
         \rowcolor{gray!11}\cite{yan2024cross,ahmed2025fedgraphmri}& Federated data augmentation & prediction model & $\bullet$ Augmented images with cross-institute knowledge & $\bullet$ tightly-coupled with the task & fastMRI and IXI\\\addlinespace
         \cite{elmas2023mri}& Federated data augmentation & shared generator & $\bullet$ decoupled knowledge with decoupled models & $\bullet$ imbalanced amount of knowledge for the generator and discriminator & fastMRI, IXI, BRATS, and in-house data\\
         \bottomrule
    \end{tabular}
    \label{tab:sum_know_exp}
\end{table*}

% Tab. of comparison
% \begin{table}
% \caption{Summary of the studies related to federated knowledge expansion.}
% \label{tab:sum_know_exp}
%     \centering
%     \begin{tabular}{l p{0.5in} p{1.4in}}
%     \toprule
%     \textbf{Branches in KAF}     & \textbf{refs}. & \textbf{pros and cons}\\
%     \midrule
%     \rowcolor{gray!7}
%     sample extrapolation   & \cite{Sanyal2019fff, FCF2019, yang2022dpmf, yang2024discrete, perifanis2022federated, JiangFedNCF2022, li2024personalized, singhal2021federated} & $\bullet$ Better privacy protection and high information density with embedding exchange\newline $\bullet$ Dependence on task-related encoders \\
%     \addlinespace
%     Data augmentation     & \cite{Kundu2024monkeypox,Ullah2024xray, Wen2022CVAE, yoon2021fedmix, Zhang2023fedm, yan2024cross, ahmed2025fedgraphmri, elmas2023mri} & $\bullet$ Expanded data size and diversified samples\newline $\bullet$ Task-independency\newline $\bullet$ Uncertain sample fidelity\\
%     \bottomrule
%     \end{tabular}
% \end{table}

\subsection{Summary}
Federated sample extrapolation mainly deals with data sparsity and missing information in local domains, whilst federated data augmentation focuses on mitigating non-IID local data distribution. We summarize related works in Table \ref{tab:sum_know_exp} and compare them in the following aspects:
\begin{itemize}
    \item \textbf{The way new knowledge is introduced}: sample extrapolation typically relies on representation learning to map multiple sources of information into one or several vector spaces, using embeddings and encoder as the knowledge carriers. By contrast, data augmentation is essentially analogous to interpolation which fuses information at sample level. This directly results in an expanded size of data and typically does not depend on any task-specific models.  
    \item \textbf{Motivation and objectives}: the extrapolation of samples is sometimes coupled with the optimization of task-specific objectives such as the accuracy of item recommendations. Federated data augmentation, in many cases, can be performed in a task-agnostic fashion.
    \item \textbf{Limitations}: For federated extrapolation, inaccurate information could be disseminated once the training samples are very noisy or manipulated by malicious clients. For augmentation, it is hard to theoretically guarantee the fidelity of synthetic samples.
\end{itemize}

%Each of the two approaches has certain limitations, as summarized in Table \ref{tab:sum_know_exp}. For collaborative sample extrapolation, domain-specific knowledge is embedded as factor matrices or vectors that are optimized under the supervision of known observations. This means that inaccurate information could be inferred and disseminated once the training samples are very noisy or manipulated by malicious clients. For collaborative data augmentation, neither interpolation-based or generative model-based schemes can theoretically guarantee the fidelity of synthetic samples. Certain mechanisms are necessary to constrain the process of augmentation against invalid patterns \cite{Jiang2024netdiff,Ktena2024gen}. So far, this problem still needs further investigation in the federated settings.

\section{Collaborative Data Refinement} % Filtering
\label{sec:filter}
Federation can bring information redundancy as well as the risk of data poisoning to the participants \cite{cina2024machine,Wang2023temporal}. We in this paper define data refinement as the techniques for excluding misinformation that exhibits knowledge incoherence. % in terms of value for learning. Knowledge filtering can involve data engineering and algorithm design at different levels across and within the domains. 

\begin{figure*}[t!]
    \centering
    \includegraphics[width=0.48\linewidth]{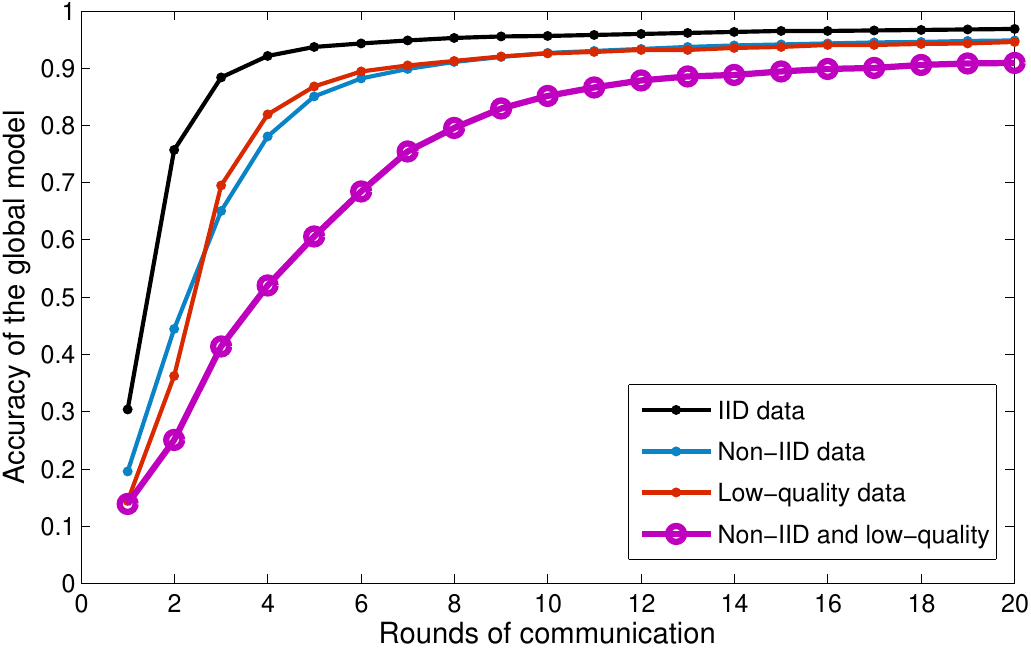}
    \includegraphics[width=0.48\linewidth]{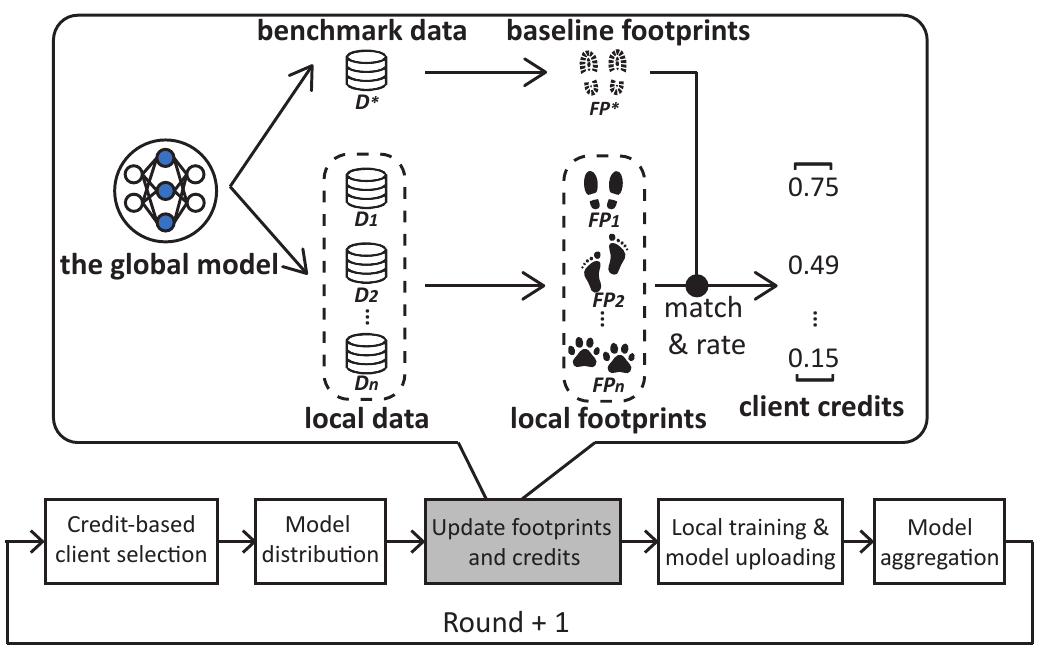}
    \caption{(Left) results of our experiments over 100 clients under four disparate conditions of data that involve IID, non-IID and low-quality on-device data. (Right) the \texttt{FedProf} framework for dataset-level filtering. Experiment details are described in \cite{wu2023fedprof}, with more observations included.}
    \label{fig:fedprof_curve}
\end{figure*}

\subsection{Dataset-level Filtering}
A coarse-grained approach to filtering out misleading knowledge is to identify and exclude the malicious participants (and their datasets). Although tailored adversarial samples are difficult to detect individually, they still inevitably change the distribution of data. Through empirical studies on label-flipping attacks, Khuu et al. \cite{Khuu2024shap} found that, with a single round of attack, a small subset of adversarial participants can result in dramatic drop in the model's performance. Inspired by the Shapley Additive Explanation (SHAP), they developed a dataset filtering algorithm using Support Vector Machines (SVMs) to classify participants based on four features derived from SHAP values. However, SHAP values are computation-intensive for each participant as we need to calculate a score matrix of all image-label pairs. By contrast, Model-based filtering approaches are low-cost and more easy-to-implement, based on the fact that poisoned model updates are likely to deviate from benign updates. For example, the \texttt{FedZZ} \cite{kumar2024precision} framework detects poisoned domains based on the cosine similarity between a client's update vector and a reference zone-aggregated vector. The proposed zone-based detection scheme groups the participants into multiple zones and calibrates the zone set progressively by accuracy. Similarly, Wei et al. \cite{Wei2024eigen} proposed to examine the eigenvalues of the covariance matrix of gradient updates. Through theoretical analysis and empirical study, they proved that gradients from adversarial participants and honest participants form clearly separable clusters, which provides forensic evidence for accurate dataset-level filtering. Li et al. \cite{li2024mitigating} proposed a confidence-aware defense (\texttt{CAD}) strategy that is insensitive to the types of poisoning. By introducing the definition of confidence score and \texttt{SuperLoss} \cite{castells2020superloss}, the rationale behind CAD is that tampered data will cause a significant drop of confidence score in local training, based on which malicious participants can be detected. %Nonetheless, challenges remain as it was also observed that the timing of attacks greatly influences the effectiveness of statistic-based defense strategies \cite{Wei2024eigen}.

Apart from adversarial participants with malicious intention, low-quality information such as noise and irrelevant samples may exist in local data. In our previous work \cite{wu2023fedprof}, we empirically investigated the impact of low-quality information on the process of collaborative learning (left panel of Fig. \ref{fig:fedprof_curve}). The result demonstrates that poor data conditions significantly slow down the convergence whilst undermining the global model's performance. We discovered that the engagement of low-quality knowledge can be as detrimental as pathological data distribution. To address the problem, a novel participant filtering strategy \texttt{FedProf} \cite{wu2023fedprof} is presented (right panel of Fig. \ref{fig:fedprof_curve}). The key aspect of this strategy is termed ``representation profiling'' where the latent output from the global model's hidden layer are extracted and compressed into participant-specific profiles. Based on the Kullback–Leibler (KL) divergence between the profiles, low-quality datasets are filtered out probabilistically before each FL round.

Noise data filtering at dataset level can be practically imperative especially for Internet of Things (IoT) applications. This is because real-time sensing is inherently susceptible to environmental disturbance and hardware flaws. For example, real-time data captured by image sensors are sensitive to violent camera shake and occlusion in the Internet of Vehicles (IoV) scenario. Targeting at this problem, Lei et al. \cite{lei2022oes} presented a filtering framework \texttt{OES-Fed} that works on the vehicle level. The framework first identifies a group of reference vehicles (i.e., vehicles with high-quality data) based on model accuracy, and sets their model parameters as the initial centers for K-means clustering. Vehicles that deviate significantly from the centers in the parameter space are detected as noise data carrier. In addition, Kalman filters and exponential smoothing are employed to incorporate the historical changes of model parameters. This enables dynamic assessment of local knowledge in response to Continual Learning (CL) scenarios \cite{liang2025diffusion} where new data keep rolling in.

\begin{table*}[htb!]
    \centering
    \caption{Summarizing the studies related to collaborative data refinement in terms of methodological highlights, drawbacks, datasets for experiments, and the form of knowledge for exchange.}
    \begin{tabular}{p{0.3in}p{0.7in}p{0.7in}p{1.6in}p{1.3in}p{1.1in}}
        \toprule
        Refs. & Category & \textbf{Knowledge carrier} & Highlights & Drawbacks & Datasets\\
        \midrule
        \rowcolor{gray!11}\cite{Khuu2024shap} & Dataset-level filtering & derived features from SHAP & $\bullet$ model-independency and statistical interpretability & $\bullet$ high computational cost of SHAP & MNIST and Fashion-MNIST\\\addlinespace
        \cite{kumar2024precision,Wei2024eigen} & Dataset-level filtering & gradients or update vectors & $\bullet$ scalable and easy-to-implement\newline$\bullet$ effective against diverse types of data pollution & $\bullet$ relatively high cost of communication \newline$\bullet$ vulnerability to collusion & Fashion-MNIST, EMNIST, CIFAR10, and LFW\\
        \rowcolor{gray!11}\cite{wu2023fedprof} & Dataset-level filtering & statistically compressed representations & $\bullet$ very low cost of communication\newline$\bullet$ stronger privacy protection& $\bullet$ requirement for dense layers in the model structure & GasTurbine, EMNIST, and CIFAR-10\\\addlinespace
        \cite{lei2022oes} & Dataset-level filtering & task-specific models & $\bullet$ scalable and effective against diverse types of data pollution\newline$\bullet$ historical behavior included for smoothing& $\bullet$ relatively high cost of communication & MNIST, CIFAR-10, and Vehicle Classification dataset\\
        \addlinespace[1pt]
        \rowcolor{gray!11}\cite{thakur2024knowledge} & Sample-level filtering & knowledge abstraction vector and the prediction model & $\bullet$ abstraction of global knowledge for local filtering\newline$\bullet$ localized filtering module adaptable to heterogeneous domains & $\bullet$ relatively high cost of communication\newline$\bullet$ vulnerability to erroneous and adversarial samples & CURIAL, eICU-CRD, and MIMIC-III\\\addlinespace
        \cite{lu2024datafree} & Sample-level filtering & models and prediction capability description & $\bullet$data-free knowledge filtering with synthetic samples\newline$\bullet$ global synthesis regulated by local knowledge ensemble & $\bullet$ high cost of communication\newline$\bullet$ vulnerability to adversaries & Fashion-MNIST, SVHN, CIFAR-10, and Digits-5\\\addlinespace
        \rowcolor{gray!11}\cite{xu2022safe} & Dataset+sample filtering & model parameters & $\bullet$ knowledge filtering at both levels using a multi-stage design\newline$\bullet$ edge-cloud hierarchical system for stronger privacy & $\bullet$ distance measure ill-suited to high dimensionality and sparsity\newline$\bullet$ applicable to DSVM only & Iris and WineQuality\\
        \addlinespace[1pt]
        \cite{tang2021data} & Federated data re-weighting & task-specific models & $\bullet$ IWDS for achieving fast convergence without loss of accuracy & $\bullet$ knowledge outside local domains unexploited & CIFAR-10 and Fashion-MNIST\\\addlinespace
        \rowcolor{gray!11}\cite{chen2024implementing} & Federated data re-sampling & task-specific models & $\bullet$ easy to implement with any classic re-sampling methods& $\bullet$ knowledge outside local domains unexploited & MIMIC-III\\\addlinespace
        \cite{ouyang2024ddpg} & Federated data re-weighting & Gini coefficients and models & $\bullet$ learnable strategy for client-specific re-weighting control\newline$\bullet$ local Gini coefficients reflecting local distributional knowledge & $\bullet$ huge state space for the DRL agent\newline$\bullet$ high communication costs & MNIST, FMNIST, and CIFAR-10\\
        \bottomrule
    \end{tabular}
    \label{tab:sum_know_filt}
\end{table*}

\subsection{Sample-level Filtering}
In contrast to the coarse-grained approaches that either keep or filter out entire local data domains, sample-level filtering identifies the deviants in each domain. 

However, it is typically not allowed in a federated setting to access the raw attributes of each individual data instances. This problem is especially intractable in the healthcare scenarios where the Electronic Health Records (EHRs) of patients are highly sensitive and probably heterogeneous across different organizations in the federation. By ``heterogeneous'' it means that samples from different clients scatter in inconsistent feature spaces that correspond to different permutations of features. To address this, Thakur et al. \cite{thakur2024knowledge} proposed a knowledge abstraction and filtering framework, where the core component is the participant-specific filtering modules that can be cascaded. Each filtering module starts with a native encoder that map local data to a unified latent space and ends with a gating operator that uses a globally shared knowledge abstraction vector to filter local samples' representations. Nonetheless, the mechanism cannot directly identify and filter out problematic samples. Motivated by the knowledge distillation paradigm \cite{le2025fedmekt,chen2023metafed}, Lu et al. \cite{lu2024datafree} proposed to employ local conditional generators to generate proxy samples for ensemble distillation. A filter is connected to the head of each local model whose outputs are then modulated based on the reference labels. This helps in refining local knowledge without the need for a public data pool.

As a matter of fact, data refinement can be performed simultaneously at dataset level and sample level. Adopting such a hybrid approach, Xu et al. \cite{xu2022safe} cast light on cloud-edge collaboration where ``dirty'' data on untrustworthy devices may significantly compromise the learning process. A novel framework called \texttt{Safe} is presented, which at the dataset level resorts to Alternating Direction Method of Multipliers (ADMM) for detecting suspicious participants that behave like data poisoning attackers. At sample level, data samples on these (potentially) malicious are then filtered based on K-means clustering and Euclidean distance. A shortcoming here is that the Euclidean distance in the raw feature space may not serve as a good criterion due to high dimensionality and sparsity. Moreover, \texttt{Safe} was only evaluated on the learning task of Support Vector machines (SVMs), which means that whether the framework can generalize to deep learning tasks remains questionable.

% % Tab. of comparison
% \begin{table}[htb!]
% \caption{Summary of the studies related to federated knowledge filtering.}
% \label{tab:sum_know_filt}
%     \centering
%     \begin{tabular}{l p{0.5in} p{1.4in}}
%     \toprule
%     \textbf{Branches in KAF}     & \textbf{refs}. & \textbf{pros and cons}\\
%     \midrule
%     \rowcolor{gray!7}
%     Client-level filtering & \cite{Khuu2024shap, kumar2024precision, Wei2024eigen, wu2023fedprof, lei2022oes} & $\bullet$ Scalable and effective against various attacks\newline $\bullet$ Strong privacy protection\newline $\bullet$ Low-sensitivity to individual sample pollution\\
%     \addlinespace
%     Sample-level filtering & \cite{thakur2024knowledge, lu2024datafree, xu2022safe} & $\bullet$ Fine-grained filtering\newline $\bullet$ Vulnerability to crafted adversarial samples\newline $\bullet$ Higher computation costs\\
%     \addlinespace
%     \rowcolor{gray!7}
%     Fed. data re-balancing & \cite{tang2021data, chen2024implementing,ouyang2024ddpg} & $\bullet$ Typically easy to implement\newline $\bullet$ Mitigation of data bias\newline $\bullet$ Reliance on class labels\\
%     \bottomrule
%     \end{tabular}
% \end{table}

\subsection{Federated Data Re-balancing} % data resampling
%The notorious problem of non-IID data across the local domains in the federation not only hampers the convergence of the global model, but also leads to the drop of accuracy when the final model is evaluated on a presumably balanced test set. From the angle of knowledge, these phenomena can be explained from two aspects. First, theoretical evidence \cite{li2019convergence,li2023nips} has revealed the convergence speed of the global model, on a certain ML task, is very much determined by the divergence of parameter updates (or accumulated gradients) from the clients. This essentially boils down to discrepancy in local knowledge as reflected by the conditional distribution of features and labels, which results in diverging local gradients towards different local optima. Second, non-IID data across the system usually implies that clients could have conflicting knowledge. Once learned and aggregated into a global model, client-specific knowledge gets diluted or neutralized, which consequently undermines the model's ability to generalize.
Data imbalance, usually associated with lopsided marginal probabilities $P(X)$ and $P(Y)$, creates difficulties for knowledge discovering. Data re-balancing has proved its effectiveness in class-imbalanced learning where re-sampling is the most common technique adopted in the literature \cite{shi2023advances,zhou2025heterogeneous}. By scaling up (down) the sampling probabilities of tail (dominant) classes, the local model is learned on a reshaped distribution of data. This can also be interpreted as a means of knowledge filtering where class-specific knowledge is kept by a different ratio. Tang et al. \cite{tang2021data} investigated the impact of the celebrated class-balanced loss (\texttt{CBLoss}) \cite{cui2019cbloss} when applied to the federated setting. Two interesting observations were made. First, data re-sampling narrowed the gap of the probability distributions of label sampling between different clients. Second, though reaching the plateau at a faster rate, re-sampling can result in an inferior model that yielded lower accuracy on local data. These two findings motivated them to devise a Imbalanced Weight Decay Sampling (IWDS) strategy that gradually smooths an imbalanced sampling towards uniform sampling. More specifically, given $n_{k,c}$ samples of class $c$ on client $u_k$, the sampling weight for a class-$c$ instance is: 
\begin{equation}
    w^{(t)}_{k,c} = \frac{1-\beta_t}{1-\beta_t^{n_{k,c}}},
    \label{eq:samp_weight}
\end{equation}
with $\beta_t < 1$ being the sampling parameter that decays from $\beta_m$ to $\beta_0$:
\begin{equation}
    \beta_t = \beta_m + (\beta_0 - \beta_m)\cdot \gamma^t,
    \label{eq:decaying_beta}
\end{equation}
where $\gamma$ is the decay rate parameter and $t$ is the index of the federated round. Using this dynamic re-sampling results in a round-varied ratio of probability that a class-$c_i$ instance gets sampled to that of a class-$c_j$ instance:
\begin{equation}
    r^{(t)}_{k, i,j} = \frac{w^{(t)}_{k,i}}{w^{(t)}_{k,j}} = \frac{1-\beta_t^{n_{k,j}}}{1-\beta_t^{n_{k,i}}}.
    \label{eq:prob_ratio}
\end{equation}

This strategy leverages imbalanced re-sampling to speed up the convergence at the early stage of collaboration and gradually approaches uniform sampling for higher final accuracy \cite{tang2021data}, i.e., $r^{(t)}_{k,i,j}$ gets closer to 1 with $\beta_t$ decreasing. Although distributional information is utilized locally to enhance local training, knowledge outside the local domain is left unexploited.

Ouyang et al. \cite{ouyang2024ddpg} formulated a different weighting rule for each sample as a power function of class frequency. After normalization the sampling probability $P_{k, c}$ for a class-$c$ instance is:
\begin{equation}
    P_{k, c} = \frac{n_{k,c}^{\beta_k}}{\sum_{i=1}^C n_{k,i}^{\beta_k}},
    \label{eq:ddpg-fl}
\end{equation}
where $C$ is total number of classes and the participant-specific parameter $\beta_k$ controls the sampling probability distribution on participant $u_k$. To dynamically determine $\{\beta_k | k=1,2,\ldots\}$ for the training cohort, a global coordinating agent based on Deep Reinforcement Learning (DRL) was developed. More specifically, the agent collects local and global model parameters as input state, makes decisions on $\{\beta_k | k=1,2,\ldots\}$, and receives rewards defined based on class-specific accuracy and Gini coefficients of local distribution.

Most of the classic re-sampling techniques for imbalanced learning can be easily applied to the local operation phase in TAF. This corresponds to a na\"ive implementation of federated data re-balancing where clients act independently based on local distribution. Chen \cite{chen2024implementing} conducted empirical studies to investigate its impacts on some downstream learning tasks. Over-sampling, under-sampling and the combination of these two were evaluated. Two interesting finds were reported: i) Under-sampling caused a slight performance loss in the final model's accuracy and precision, and ii) the hybrid method (combining over-sampling and under-sampling) showed the best performance across the board in all metrics. These observations imply a strong potential of data re-balancing in the federation even without any knowledge exchange between the domains. However, whether the phenomenon still holds for more complex tasks remains under-explored.

We summarize these solutions in Table \ref{tab:sum_know_filt}.

\subsection{Summary}
With rising concerns of data-centric attacks, dataset-level filtering can minimize the risk of involving poisonous data. We argue that fine-grained filtering at sample level can be employed jointly with participant screening to provide more reliable evidence for adversary detection. As of data re-balancing, existing approaches stand on an implicit assumption that uniform data distribution is ideal for the downstream task. However, this may not always hold. On the one hand, the local learner could under-fit on the ``special knowledge'' \cite{tang2021data}. On the other hand, there is no sufficient theoretical evidence that supports imbalanced sampling, provided that the test data distribution is unknown. Besides, collaborative data refinement for regression tasks, e.g. drug response modeling, is hardly explored. Moreover, existing studies pay little attention to the subtle line between data imbalance and non-IIDness of data, the understanding of which can help us better redefine how data can refined locally and globally. %In Table \ref{tab:sum_know_filt} we summarize the strength and limits of existing studies in relevance to knowledge filtering in the federation.

\section{Collective Harmonization of Data} % Modification
\label{sec:correct}
This section revisits branches of studies that employ collaborative mechanisms to make correction or transformation on data labels or features so that they can be utilized later in a harmonized fashion. %In contrast to knowledge filtering that aims to exclude misinformation, amendments of samples usually involves data engineering techniques and feature engineering methods. Hence, in the following content we survey the state-of-the-art under two different categories: federated data correction and federated feature engineering.

\subsection{Federated Data Correction}
Mislabeled samples can have salient impact. This motivates researches on federated noisy label learning (FNLL) \cite{jiang2024tackling,wu2023fednoro}. Considering the tie between a participant and its local domain, Xu et al. \cite{xu2022fedcorr} proposed \texttt{FedCorr}, a multi-stage label denoising framework that streamlines a pre-processing stage and a fine-tuning stage prior to the traditional learning process, all executed in a federated fashion. In the pre-processing stage, each participant computes an estimate of the Local Intrinsic Dimensionality (LID) based on its local model's predictions, which are collected by the central server for noisy dataset detection. As an useful measure that connects dimensionality and discriminability \cite{houle2013dimensionality,karger2002finding}, the LID for a vector $\boldsymbol{x} \sim\mathcal{D}$ is defined as:
\begin{equation}
    {LID}_x = \lim_{r\rightarrow 0} 
        \lim_{\epsilon \rightarrow 0} \frac{\log F_{d_x}((1+\epsilon)r) - \log F_{d_x}(r)}{\log (1+\epsilon)},
    \label{eq:LID}
\end{equation}
where $d_x$ is the distance between $\boldsymbol{x}$ and a randomly drawn vector $\boldsymbol{x}'$ from $\mathcal{D}$ and $F_{d_x}(\cdot)$ represents the cumulative distribution function (CDF) of $d_x$. As models trained on noisy data tend to report larger LID scores, the server then uses this rule to spot these participants for label correction. 

\texttt{FedCorr} \cite{xu2022fedcorr} adopts a confidence-based rule to find and correct mislabeled data on noisy datasets. Specifically, the subset of samples (controlled by a proportion $\pi$) that yield the highest losses is first selected, following which the predictions of the global model $f_{G}$ are used for relabeling. Formally, the final subset of samples $D'_k$ (for a noisy client $u_k$) to be relabeled is defined in (\ref{eq:fedcorr_subset}):
\begin{multline}
    D'_k = \big\{ (\boldsymbol{x},y) | (\boldsymbol{x},y) \in \\ \underset{\hat{D}\subseteq D_k \atop |\hat{D}|=\pi|D_k|}{\arg\max} \sum_{(\boldsymbol{x},y)\in \hat{D}} \ell(f_{G}(\boldsymbol{x}),y) \text{ and } \max f_{G}(\boldsymbol{x}) \geq \theta \big\},
    \label{eq:fedcorr_subset}
\end{multline}
where $\ell$ is the loss function and $D_k$ is the local data from participant $u_k$. The labels of $D'_k$ are then corrected to the predictions by the global model $f_{G}$.

There are two major defects in \texttt{FedCorr} \cite{xu2022fedcorr}: i) incompatibility to general loss functions other than cross entropy, and ii) susceptibility to dishonest participants who tamper with their LID scores. Targeting at the second issue, Wang et al. \cite{wang2024icde} built upon \texttt{FedCorr} and redesigned its federated pre-processing stage to achieve robust label correction. The key idea behind is to craft an efficient Zero-Knowledge Proof (ZKP) \cite{campanelli2022zkp} protocol by which the server (as the verifier) can verify the computation integrity of the LID scores reported by the participants (as the provers) in a batched manner. The proofs are generated both before and after label correction. Only calculation in the form of arithmetic circuits is verifiable by the protocol, which limits its potential application to other data correction schemes. Similarly, the \texttt{FedELC} framework by Jiang et al. \cite{jiang2024tackling} detects participants that carry noisy data and corrects them in two separate stages. The first stage is almost the same as in \cite{xu2022fedcorr,wu2023fednoro}, whilst the second stage creates learnable soft labels for data correction optimized towards the minimization of a linear combination of three different entropy losses. \texttt{FedELC} was evaluated on a rich set of noise data settings including the CIFAR-N \cite{wei2021cifar-n} dataset from Amazon Mechanical Turk which features a high noise rate of human-annotated labels.

Noisy label detection and correction can also be executed directly at sample level. Based on label uncertainty estimation \cite{northcutt2021confident}, Zeng et al. \cite{zeng2022clc} started from utilizing a well-trained global model's predictions to estimate the probability $P_{k,i}$ that samples of a class $i\in \mathcal{C}$ are correctly labeled in a local dataset $D_k$, given in (\ref{eq:clc_local}):
\begin{align}
    P_{k,i} = & \sum_{j\in \mathcal{C}} p(y=i|y^*=j) p(y^*=j|y=i) \nonumber \\
        \approx &  \sum_{j\in \mathcal{C}} p(y=i|y^*=j) \mathbb{E}_{\boldsymbol{x}\in D_{k,y=i} }p(y^*=j;\boldsymbol{x}, \theta_G) \nonumber \\
        = & \mathbb{E}_{\boldsymbol{x}\in D_{k,y=i}}\sum_{j\in \mathcal{C}} p(y=i|y^*=j;\boldsymbol{x}, \theta_G) p(y^*=j;\boldsymbol{x}, \theta_G) \nonumber \\
        = & \mathbb{E}_{\boldsymbol{x}\in D_{k,y=i}}\sum_{j\in \mathcal{C}} p(y=i, y^*=j;\boldsymbol{x}, \theta_G) \nonumber \\
        = & \mathbb{E}_{\boldsymbol{x}\in D_{k,y=i}}p(y=i;\boldsymbol{x}, \theta_G) \nonumber \\
        = & \frac{1}{|D_{k,y=i}|}\sum_{v=1}^{|D_{k,y=i}|} p(y_v=i;\boldsymbol{x}_v, \theta_G),
    \label{eq:clc_local}
\end{align}
where $D_{k,y=i}$ is the data subset on participant $u_k$ labeled as class $i$, $\mathcal{C}\subset \mathbb{W}$ is the label space, and $\theta_G$ denotes the global model's parameter. Next, the global estimate of the probability of correctly-labeled class $i$ can be obtained using weighted average consensus \cite{zeng2022clc} over the population $\mathcal{K}$: 
\begin{equation}
    P_{i} = \sum_{k\in \mathcal{K}} \frac{|D_{k,y=i}|}{\sum_{k\in \mathcal{K}}|D_{k,y=i}|} P_{k,i},
    \label{eq:clc_global}
\end{equation}
which resembles the model aggregation rule in the canonical \texttt{FedAvg} \cite{mcmahanfl}.

A practical concern of federated data correction is that the participants can be self-interested, which means that they may not stick to the protocol if their gain of participation cannot cover the cost of data correction. In light of this issue, Yan et al. \cite{yan2024incent} first modeled the payoff of participants as a function of the model's accuracy, payment from the task publisher, and the data correction cost. The authors employed an iterative denoising strategy that corrects a fraction of $\alpha$ noise samples per round. To reduce the fraction of noisy data from $\epsilon$ to $\epsilon'$ given a denoising cost per sample $c$ and data size $|D_k|$, the total cost of data correction for client $u_k$ is:
\begin{equation}
    {Cost}_k = |D_k| c (\log_\alpha \epsilon - \log_\alpha \epsilon').
\end{equation}

Yan et al. \cite{yan2024incent} considered two different situations of federated denoising. In the first case, participants are cooperative and work towards to the optimal social welfare, i.e., the maximization of the sum of client payoffs. In the second case, every participant is assumed self-interested and only motivated to maximize its own payoff, which corresponds to a non-cooperative game.

\subsection{Federated Feature Engineering} % selection, transformation
\begin{figure}[tb]
    \centering
    \includegraphics[width=0.99\linewidth]{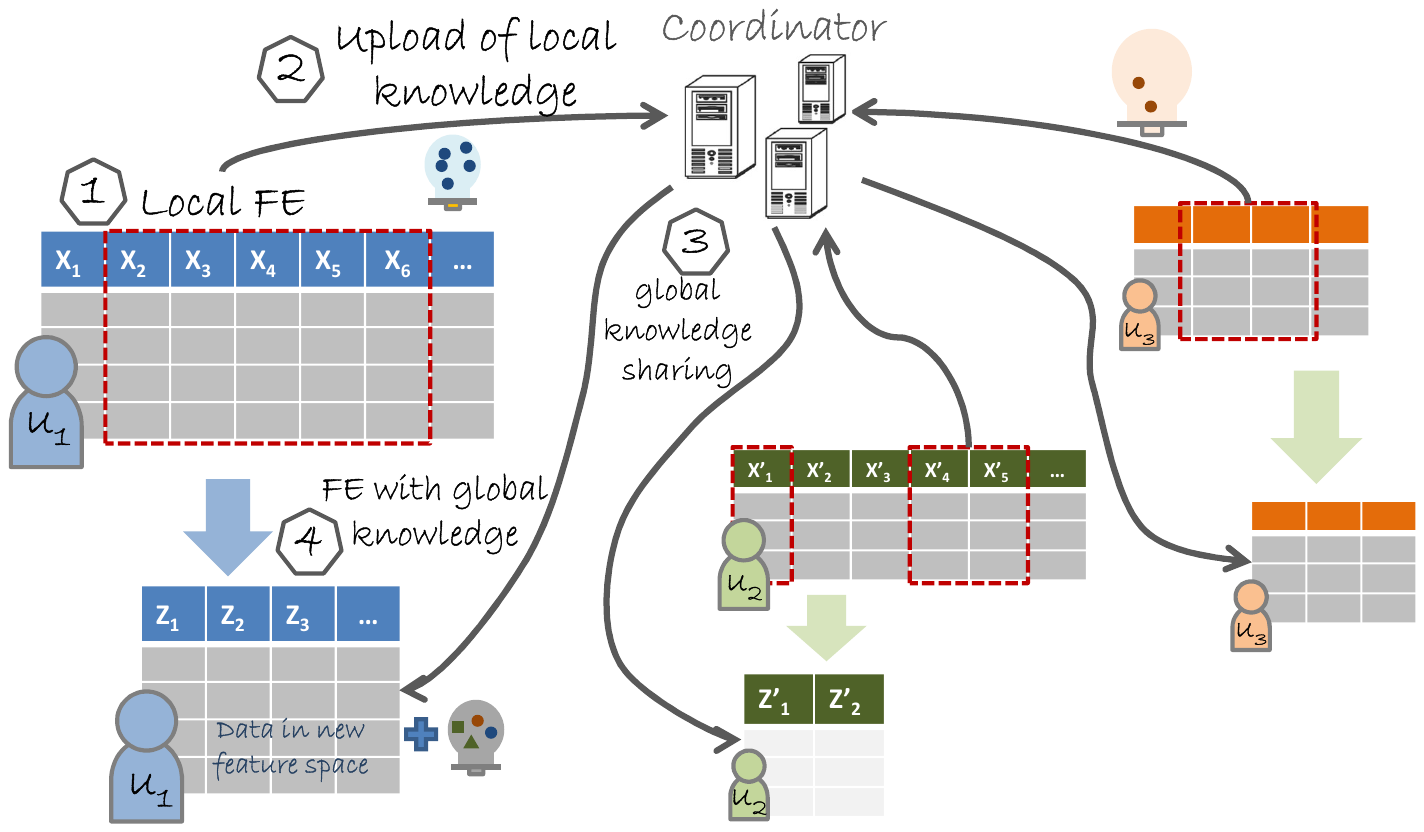}
    \caption{A schematic illustration of federated feature engineering.}
    \label{fig:fedfe}
\end{figure}

In TAF, by federated feature engineering we refer to a new thread of research with particular focus on aligning, selecting and transforming the feature space (Fig. \ref{fig:fedfe}). This process can either be independent of the downstream task or somehow coupled with a specific protocol, motivated to address three challenges: 

\begin{itemize}
    \item \textit{Disordered or misaligned attributes}: in cross-institutional cooperation it is common to deal with different views of data where the local datasets are subject to different permutations of attributes \cite{thakur2024knowledge}.
    \item \textit{Heterogeneous feature space}: from a single domain to a multitude of domains with partial overlap in the feature space, the heterogeneity brings great difficulty to the design of autonomic feature engineering.
    \item \textit{Varied significance of features}: non-IID settings usually mean bias and special knowledge within each domain, which will greatly influence the importance of features when measured statistically.
\end{itemize}

These challenges have catalyzed cutting-edge studies on new feature engineering paradigms. In this section we discuss the advance in federated feature engineering, with particular focuses on collaborative feature selection and feature transformation.

\texttt{Fed-FiS} \cite{banerjee2021fedfis} represents an early exploration of collaborative feature selection in the federation based on information theory. The study follows a training-free, task-agnostic approach to assessing the importance of features. Locally, a participant $u_k$ independently evaluates the feature-feature mutual information (FFMI) and the feature-class mutual information (FCMI) over the local feature and label set $\{f^{(k)}_1, f^{(k)}_2, \ldots\} \cup C^{(k)}$ by Eqs. (\ref{eq:ffmi}) and (\ref{eq:ffci}):
\begin{equation}
    {FFMI}^{k}_{i,j} = \sum_{x\in f^{(k)}_i} \sum_{y\in f^{(k)}_j} p(x,y) \log \frac{p(x,y)}{p(x)p(y)},
    \label{eq:ffmi}
\end{equation}
\begin{equation}
    {FFCI}^{k}_i = \sum_{x\in f^{(k)}_i} \sum_{c\in C^{(k)}} p(x,c) \log \frac{p(x,c)}{p(x)p(c)}.
    \label{eq:ffci}
\end{equation}

Based on these mutual information-based measures, each participant performs clustering to select a subset of features which are then uploaded to the central server. In the second stage, the server uses FFMI and FCMI again to score each feature, which in turn helps the participants identify the best local feature subset. Banerjee et al. \cite{banerjee2024cost} further extended this framework using an improved ranking scheme based on Pareto optimization where each feature is treated as a solution characterized by FFMI and FCMI. These solutions are then ranked based on how many others dominate them and that they dominate. Rankings are reported back to the clients. 

Fu et al. \cite{fu2023feast} studied the feature selection problem in the Vertical Federated Learning (VFL) setting where the set of entities are associated with consistent identifiers but represented by different features. The proposed framework, \texttt{FEAST}, involves three types of participants for collaborative feature selection: a selected party, an active party, and candidate parties. In the pre-processing stage, discretization is first applied to each feature, combined with stratified sampling to reduce computation and communication costs. This is followed by iterative operations and interactions between the parties. Specifically, a party is selected to report its statistical variables which are generated by grouping local features and re-encoding the instances. The candidate parties receive the statistical variables and score each feature based on Conditional Mutual Information (CMI) \cite{yang1999feature}. The feature scores are then collected by an active party to perform global ranking to produce the best feature subset $S_k$ (s.t., $|S_k|=m$ given $m$ as the number of local features to select) for each participant $u_k$, which collectively represents the solution to the following optimization: 
\begin{equation}
    \min_{S_k\subseteq \mathcal{X}_k} H(Y|\{S_1, S_2,\ldots\}),
\end{equation}
where $H(Y|\{S_1, S_2,\ldots\})$ is the conditional entropy of the label $Y$ given the union of locally selected feature subsets $\{S_1, S_2,\ldots\}$. This framework offers useful insights for communication-efficient exchange of knowledge for collaborative feature selection, but the protocol does not consider dishonest participants in the system. As a matter of fact, feature engineering techniques can be utilized to defend against data poisoning attacks strengthened by feature manipulation. This is also termed feature poisoning (FP) in the literature \cite{Nowroozi2024rdfs}. Experimental evidence has demonstrated that, despite being highly disguised, FP attacks can be effectively countered by the Random Deep Feature Selection (RDFS) \cite{Nowroozi2025fp} strategy. 

Beside feature selection, feature transformation represents another typical approach to data engineering. By mapping samples to a new space of features derived from raw data, feature transformation not only helps to reduce information redundancy but also refines the representation of knowledge that is difficult to learn. However, it is not trivial for data holders in a federated system to perform feature transformation collaboratively. A practical problem that stands out is the computation cost incurred by performance evaluation given numerous combinations of transformed features. Moreover, privacy issues arise since transformation can involve multiple features coming from different domains, which necessitates Secure Multi-Party Computation (MPC). To address these issues, Fang et al. \cite{fang2020flfe} designed \texttt{FLFE}, a federated feature engineering framework for efficient, privacy-preserving multi-party collaboration. The core techniques employed include a sketching scheme for quantizing local features, a DNN-based classifier for judging the value of features, and a mask-based feature exchange protocol for transformation. This brings advantages in communication efficiency and also preserves data privacy without homomorphic encryption that limits the diversity of transformations.

\begin{table*}[htb!]
    \centering
    \caption{Summarizing the studies related to collective harmonization of data in terms of methodological highlights, drawbacks, datasets for experiments, and the form of knowledge for exchange.}
    \begin{tabular}{p{0.3in}p{0.8in}p{1.0in}p{1.6in}p{1.1in}p{1.1in}}
        \toprule
        Refs. & Category & \textbf{Knowledge carrier} & Highlights & Drawbacks & Datasets\\
        \midrule
        \rowcolor{gray!11}\cite{xu2022fedcorr,wang2024icde} & Federated label correction & local intrinsic dimensionality (LID) and models & $\bullet$ global pre-processing plus ``clean'' clients-only fine-tuning\newline $\bullet$ local knowledge encapsulated in LID & $\bullet$ incompatibility to general loss functions & CIFAR-10, CIFAR-100, Clothing1M, and simulated data \\
        \cite{jiang2024tackling} & Federated label correction & warm-up models and class-wise loss vectors & $\bullet$ distance-aware aggregation to mitigate noisy knowledge\newline$\bullet$ learnable soft labels for local correction & $\bullet$ sensitive to the setting of multiple hyper-parameters & CIFAR-10/100, CIFAR-10/100-N, Clothing1M \\
        \rowcolor{gray!11}\cite{zeng2022clc} & Federated label correction & class-wise confidence thresholds and models & $\bullet$ consensus-based method to ensemble decentralized knowledge &$\bullet$ reliance on a pre-trained model for label uncertainty estimation & MNIST, USC-HAD, CIFAR-10, and Clothing1M\\
        \cite{yan2024incent} & Federated label correction & denoising rate decisions and models &$\bullet$ payoff modeling for game-based interaction\newline$\bullet$ iterative denoising with a tunable fraction & $\bullet$ reliance on accurate estimate of noise rate of private data & MNIST and CIFAR-10\\
        \addlinespace[1pt]
        \rowcolor{gray!11}\cite{banerjee2021fedfis,banerjee2024cost} & Federated feature selection & MI-based scores and feature subsets& $\bullet$ low communication costs\newline$\bullet$ model-agnostic and task-independent & $\bullet$ lack of personalization\newline$\bullet$ applicable to single-label tasks only & SL-KDD99 and anonymized credit card transactions (ACC)\\\addlinespace
        \cite{fu2023feast} & Federated feature selection & statistical variables and feature scores & $\bullet$ MI-based and model-independent\newline$\bullet$ efficient exchange of statistical knowledge & $\bullet$ vulnerability to dishonest parties\newline$\bullet$ limited to VFL on tabular data & MIMIC III, PhysioNet, Census-Income, and Nomao\\\addlinespace
        \rowcolor{gray!11}\cite{zhang2023fshfl} & Federated feature selection & feature subsets & $\bullet$ low communication cost\newline$\bullet$ unsupervised and model-agnostic feature selection & $\bullet$ training required for each local OCSVM\newline$\bullet$ lack of personalization & Bot-IoT, ACC, KDD99, and DEFT\\
        \cite{cassara2022tvt} & Federated feature selection & probability vectors associated with features & $\bullet$ probabilistic feature subset selection and Bayesian aggregation\newline$\bullet$ significantly reduced network traffic & $\bullet$ convergence guarantee for IID data only & MAV and WEarable Stress and Affect Detection (WESAD) \\
        \addlinespace[1pt]
        \rowcolor{gray!11}\cite{overman2024federated} & Federated feature transform & expression of derived features & $\bullet$ low communication cost and reduced search space & $\bullet$ very high cost of feature subset evaluation & OpenML586 and Airfoil\\\addlinespace
        \cite{fang2020flfe} & Federated feature transform & sketches of features & $\bullet$ Quantile Data Sketch and mask-based feature exchange\newline$\bullet$ model-free evaluation & $\bullet$ pre-training required for the DNN-based feature judge & 166 datasets from OpenML\\        
         \bottomrule
    \end{tabular}
    \label{tab:sum_correct}
\end{table*}

A major obstacle for adapting Automated Feature Engineering (AutoFE) methods (e.g., OpenFE \cite{zhang2023openfe}) to the federated system is the prohibitive cost of communication. To this end, Overman and Klabjan \cite{overman2024federated} have made early effort and devised a federated framework for collaborative AutoFE. In their framework, participants independently carry out AutoFE and report the expression of derived features to the server. The server then randomly applies masks to the union of these features to generate feature subsets. An iterative halving strategy is used to find the best mask (e.g., optimal feature subset) after running multiple rounds of learning for performance evaluation. For participants with heterogeneous feature space, feature values across the entire local datasets need to be shared after homomorphic encryption so as to fuse features from different domains. For hybrid settings, importance of common features across multiple domains can be assessed progressively with the process. 

Federated feature engineering can facilitate the refinement of knowledge in a broad range of scenarios. For example, Zhang et al. \cite{zhang2023fshfl} proposed a federated feature selection framework for Internet of Things (IoT) applications where each device (as a participant) has very limited computational resources. In this framework, local features $\{f^{(k)}_1,f^{(k)}_2,\ldots\}$ on each client $u_k$ are first filtered using a threshold defined based on Symmetric Uncertainty (SU) and Feature Average Relevance (FAR):
\begin{equation}
    {SU}(f^{(k)}_i, f^{(k)}_j) = 2\Big[ \frac{{MI}(f^{(k)}_i,f^{(k)}_j)}{H(f^{(k)}_i)+H(f^{(k)}_j)} \Big],
\end{equation}
\begin{equation}
    {FAR}(f^{(k)}_i) = \frac{1}{d_k}\sum_{j=1}^{d_k} {SU}(f^{(k)}_i, f^{(k)}_j),
\end{equation}
where $d_k$ is the feature dimensionality for participant $u_k$, $H(\cdot)$ is the entropy measure and $MI(\cdot, \cdot)$ denotes the mutual information between two features. Next, One-Class Support Vector Machine (OCSVM) and hierarchical clustering are employed on each participant to identify the subset of most significant features. The server then calculates the intersection of these subsets and synchronizes with the participants. The solution was evaluated on four classification tasks in relevance to IoT applications. Experimental results demonstrated that competitive and even higher performance could be achieved with the federated feature subset, with improved efficiency.

Cassar\'{a} et al. \cite{cassara2022tvt} argued that a small subset of features are adequate for an interconnected Autonomous Driving Systems (ADSs). To this end, they developed a federated feature selection algorithm that combines on-vehicle feature filtering and edge-based feature subset aggregation. Locally, each participating vehicle optimizes a feature selection probability vector such that the conditional entropy $H(y|S)$ (given feature subset $S$ and labels $y$) is minimized, which corresponds to an Associated Stochastic Problem (ASP). The edge server then aggregates these probability vectors into a consensus one and feeds it back to the vehicles. This interaction keeps going iteratively until the global vector converges. The Kolmogorov-Smirnov test with preset parameters are used for checking convergence. By evaluating this algorithm, the authors claimed that a 99\% compression rate for network traffic was achieved since as few as 24 (out of 2166) features were selected for collaborative training on the MAV dataset, with negligible impact on downstream tasks. This finding indicates that federated feature engineering can be as effective as model-based compression methods (e.g., quantization, pruning, etc.) \cite{gupta2022compression} in terms of cost saving.

% % Tab. of comparison
% \begin{table}[htb!]
% \caption{Summary of the studies related to federated label and feature space correction.}
% \label{tab:sum_correct}
%     \centering
%     \begin{tabular}{l p{0.5in} p{1.3in}}
%     \toprule
%     \textbf{Branches in KAF}     & \textbf{refs}. & \textbf{pros and cons}\\
%     \midrule
%     \rowcolor{gray!7}
%     Fed. data correction & \cite{xu2022fedcorr,wang2024icde, jiang2024tackling, zeng2022clc, yan2024incent} & $\bullet$ Effective against noise and adversaries\newline $\bullet$ Reliance on well-trained models and labels\\
%     \addlinespace
%     Fed. feature engineering & \cite{banerjee2021fedfis, banerjee2024cost, fu2023feast, zhang2023fshfl, cassara2022tvt, overman2024federated, fang2020flfe} & $\bullet$ Generally independent of models and tasks\newline $\bullet$ Good protection of data privacy\newline $\bullet$ Susceptible to ``curse of dimensionality''\\
%     \bottomrule
%     \end{tabular}
% \end{table}

\subsection{Summary}
In this section we highlight emerging studies that tailor the dimensions in which data are represented. Federated data correction attempts to amend the misinformation carried by the data to avoid erroneous knowledge, with particular focus on the labels. Reference knowledge, in most of the label-correction methods, is encapsulated in a global model. This limits their application to task-agnostic scenarios. Federated feature engineering focuses on identifying the most important attributes or the most expressive combination of them. In this context, the statistical properties of features are often utilized as the measure of knowledge they can provide. Their pros and cons are summarized in Table \ref{tab:sum_correct}.

Although being different in methodologies, the two lines of research have much in common. First, both data correction and feature engineering are typically implemented as part of a multi-stage procedure in the federation. For example, data correction is usually cooperated by the participants in the pre-processing stage before the downstream tasks. Similarly, federated feature engineering usually involves repetitive local feature assessment and client-server information exchange. 
In addition, we discover that both approaches can be sensitive to the discrepancy of data distribution across different domains. A unified posterior (that describes the conditional probability of labels) or an optimal feature subspace may not work well for every individual participant in the federated system.

\section{Open Challenges and Questions}
Through revisiting diverse approaches which bring generic knowledge into local data domains in a federated system, we argue that there is a large space ahead to explore especially in the aspects of efficiency and efficacy. In this section we highlight opportunities for future research in relevance to TAF. %The advance in knowledge discovery combined with the promising collaborative learning is expected to open up new opportunities for data-centric, learning-driven collaboration in the future. %In the paper we highlight urgent challenges and open questions that involve but are not limited to the ways knowledge is learned, represented and exchanged as well as how to enhance Large Language Models (LLMs) with KAF. We expect to offer unique insights for research beyond knowledge for learning.

\subsection{Decentralized Peer-to-Peer Knowledge Seeking}
%Extensive effort has been made in decentralized machine learning since it has natural advantages in scalability, flexibility and byzantine-robustness \cite{vincent2023systematic,moussa2023towards}. However, existing approaches are mostly learning-centric, which means that their objectives are shaped towards the optimization of a specific group of models or a family of relevant tasks. This makes the collaboration costly and hard to reproduce, because the model as the knowledge carrier is very sensitive to the uncertainty in federated training. This uncertainty boils down to the population involved as well as the sequence of knowledge propagation in the network, which also raises the cost of reproducing the model. 
%Recent studies on decentralized matrix factorization (DMF) present a domain-specific path to expand local knowledge domain via peer-to-peer (P2P) communications \cite{hou2023bfrecsys,huang2024towards}. DMF can be regarded as a special case of decentralized knowledge augmentation. 
Decentralization has natural advantages in scalability, flexibility and byzantine-robustness \cite{vincent2023systematic}. Recent studies on decentralized matrix factorization (DMF) present a domain-specific path to expand local data via peer-to-peer (P2P) communications \cite{hou2023bfrecsys,huang2024towards}. DMF can be regarded as a special case of decentralized data augmentation, yet it is hard to deal with general forms of data other than matrices. With a novel vision, Abdelmoniem et al. \cite{abdelmoniem2025} described a decentralized knowledge routing framework where a participant can describe the knowledge requested and find another well-matched peer for collaborative learning. This is done by sharing artificial samples with the proxy devices (called knowledge discoverers) responsible for request routing and handling. Though the objective employed is still task-specific and learning-oriented, it inspires us to explore new directions for decentralized knowledge exchange and knowledge matching between domains.
\par\noindent\emph{Clues for further study: knowledge abstraction and matching \cite{wu2023fedprof, ahmed2024fi}, blockchain, zero-knowledge proof, and knowledge routing \cite{abdelmoniem2025}.}

\subsection{Light-weight Knowledge Courier and Efficient Exchange}
A key factor in task-agnostic federation is the exchange of knowledge between distributed local domains. For example, the item factor matrix is shared as the carrier of domain-specific knowledge in federated collaborative filtering, whilst the global model takes this role in federated label denoising. Whatever form it takes for knowledge exchange, the reduction of communication cost persists as a top priority in practice. 

On the application layer, the roadmap for efficient knowledge exchange mainly encompasses two approaches: i) decreasing the communication payload incurred by the courier of knowledge, and ii) reducing the number of communication rounds if iterative interactions are needed. Existing studies often exchange models, but it is worth noting that models, model updates and gradients are ``heavy''. Provided that it is not necessary to exchange the ``full picture'' of knowledge in many situations such as federated data filtering and harmonization, how to craft a flexible protocol to avoid information redundancy emerges as new problem to be resolved. 
\par\noindent\emph{Clues for further study: model-free knowledge representation, information bottleneck \cite{yang2024ib}, and Bayesian compression \cite{yang2025continual}.}

\subsection{Federated Retrieval of Exogenous Knowledge with Confidentiality} % RAG
Latest LLM-empowered dialogue systems \cite{guo2025deepseek,openaio1systemcard} commonly adopt Retrieval-Augmented Generation (RAG) to bring in up-to-date, verified information so as to produce accurate knowledge, a.k.a. reducing hallucinations. With rising awareness of intellectual properties, recent studies have paid attention to the federated Retrieval-Augmented Generation (FedRAG) \cite{shojaee2025federated} with requirements of context confidentiality \cite{addison2024cfedrag}. This makes a special case of TAF where knowledge exchange (including queries and context) happens between multiple knowledge bases (KBs) and a reasoning agent. In this scenario, few studies stand on the KB side and they hardly account for the cost of communication and the price of retrieval service. On the one hand, it is very expensive in time to reliably determine whether a huge database contains the piece of knowledge desired. On the other hand, the scarcity of user feedback brings difficulties for assessing the usefulness of contents retrieved. In addition, it is worthwhile to explore how multiple knowledge bases can cooperate with each other, e.g., in a TAF setting, to provide refined and logically consolidated retrieval capability.
\par\noindent\emph{Clues for further study: federated RAG with confidentiality \cite{addison2024cfedrag}, federated search systems \cite{guerraoui2025efficient}, query-content evaluation \cite{yu2024knowledge}, and scattered knowledge bases \cite{wu2025multirag}.}

\subsection{Task-agnostic Collaboration over Incomplete Data}
We observed that the majority of researches in the federated setting assume the completeness of data, overlooking the fact that real-world data have missing attributes and labels. %This is mainly attributed to the prior knowledge implied by the labels that usually serve as the most important criterion for many operations. 
Especially, many existing studies rely on data labels for knowledge extraction and empirically associate non-IIDness to the marginal distribution of labels. For example, feature engineering typically relies on feature-label mutual information, and data re-balancing strategies re-weight data points by class label. This means that they cannot apply to collaboration over incomplete data. This largely limits their application to a broader categories of unlabeled data such as text corpora, surveillance footage and browsing records. We argue that it is important while challenging for the data owners in TAF to represent incomplete data in a traditional form. For example, it is impossible to swap class-wise information over unlabeled datasets. In this case, we suggest further exploration into flexible representations of data based on their intrinsic structures and robust protocols for knowledge exchange between parties of varied data conditions.
\par\noindent\emph{Clues for further study: fuzzy representations, auxiliary supervision \cite{yang2024depth}, sample fusion, and knowledge-data alignment \cite{zhao2024weakly} .}

\subsection{Can TAF Make Autonomy Superior over FL?}
\textit{We argue that the gain of collaborative learning becomes marginal when the participants' data get larger and diversified.} This is the key factor that limits the sustainability of learning-centric collaboration such as FL. On the contrary, task-agnostic federation liberates the participants from having to work on a specific model. This creates possibility for each participant to learn independently after the collaboration. But how competitive will it be regarding the downstream task when compared to federated learning? \emph{The answer is uncertain} and hinges on multiple factors including local data size, the test data distribution, and the utility of data gained via TAF. In terms of performance, TAF can increase the size and diversity of samples necessary for independent training, with augmented knowledge from foreign domains for generalizable model. In terms of cost, FL struggles in communication efficiency with large models and huge participant population, whilst TAF plus independent training has tempting potential when equipped with efficient knowledge exchange protocols. From both perspectives, it is still an open-ended question.
\par\noindent\emph{Clues for further study: benchmarking, query-based knowledge transfer \cite{alballa2025qkt}, and autonomous organization \cite{zhang2025srfl}.}

\subsection{Can TAF Support Knowledge Erasure?}
``The right to be forgotten'', as stated in privacy-protecting regulations, has motivated recent studies in unlearning knowledge gained in a group \cite{chen2024fmu}. However, many existing unlearning approaches sit on unrealistic assumptions such as storing the entire trace of gradients or forcing all participants to engage in the unlearning stage \cite{tarun2024fast}. More importantly, they are yet to be evaluated on realistic unlearning scenarios where the shared knowledge needs to be ``withdrawn'' any time on demand. In contrast to these traditional settings, TAF opens up new opportunities for knowledge erasure. On the one hand, collaboratively augmented data can be reckoned as ``borrowed'' knowledge which is easy to remove. On the other hand, we believe it is also possible to accelerate any downstream unlearning task by exchanging artificial samples dedicated for ``overwriting'' specific knowledge. In both ways TAF can support efficient knowledge erasure. It also calls for innovative data engineering and knowledge exchange protocols in response to this special demand.
\par\noindent\emph{Clues for further study: Augmentation-based unlearning \cite{falcao2025data,Choi2024CVPR}, Adversarial Mixup \cite{peng2025adversarialmixup}, inhibited synthetic data \cite{zhang2025toward}, influence-aware filtration \cite{dang2025efficient}, and noise learning \cite{chien2024langevin}.}

\section{Conclusion}
As the value of data assets and privacy issues arise, it is of utmost importance for data owners to make full use of proprietary information whilst seeking cooperation with their counterparts. Although federated learning-based paradigms have paved the way for privacy-preserving collaboration, its learning-centric nature greatly limits its feasibility over cost-sensitive, self-interested data owners. We argue that it is worth thinking of a new scenario by shifting the focus to data-centric and task-irrelevant cooperation, which substantially diversifies the way how participants can benefit from the federation. With this vision, we first conceptualize a new scenario of collaboration termed Task-Agnostic Federation, and then investigate latest research that enables exchange of generic knowledge to achieve collaborative data expansion, data refinement and data harmonization over decentralized data. Through a comprehensive literature review, we point out that the carrier of knowledge and the protocol for knowledge exchange are the most influential factors for TAF to be effective, efficient and extendable. Based on our investigation, we further highlight a number of open challenges that are not yet adequately addressed by existing effort, and also pose two open-ended questions to inspire more theoretical and empirical exploration in the domain.

\section*{Acknowledgement}
This work is supported by the National Natural Science Foundation of China (62402198, U23B2027), National Social Science Foundation of China Post-funded Project (20FGLB034), Basic and Applied Basic Research Project of Guangzhou (2025A04J2212), the Fundamental Research Funds for the Central Universities (21624348), and New Generation Artificial Intelligence-National Science and Technology Major Project (2025ZD0123605).

An earlier preprint version of this article entitled ``Knowledge Augmentation in Federation: Rethinking What Collaborative Learning Can Bring Back to Decentralized Data'' was posted on arXiv (DOI arXiv:2503.03140).

\ifCLASSOPTIONcaptionsoff
  \newpage
\fi



%\bibliographystyle{IEEEtran}
%\begin{thebibliography}{1}


%\end{thebibliography}

\bibliography{refs}
\bibliographystyle{unsrt}

\end{document}